\documentclass[hyper]{JHEP3}

\usepackage{amsmath,amsthm}
\def\be{ \begin{equation} }
\def\ee{ \end{equation}}

\def\Aut{{\rm Aut}}

\def\Co0{{\rm Co}_0}

\def\dim{{\rm dim}}
\def\exp{{\rm exp}}

\def\Fix{{\rm Fix}}

\def\I{{\rm i}}

\def\log{{\rm log}}
\def\mod{{\rm mod}}

\def\sym{{\rm Sym}}

\def\Tr{{\rm Tr}}

\def\half{\frac{1}{2}}

\def\one{{\hbox{ 1\kern-.8mm l}}}

\def\CC {{\cal C}}
\def\CD {{\cal D}}

\def\CF {{\cal F}}
\def\CG {{\cal G}}
\def\CH {{\cal H}}

\def\CL {{\cal L}}
\def\CM {{\cal M}}
\def\CN {{\cal N}}
\def\CO {{\cal O}}

\def\CR {{\cal R}}
\def\CV {{\cal V}}

\def\CO {{\cal O}}

\def\CG {{\cal G}}
\def\CH {{\cal H}}

\def\IF{\mathbb{F}}

\def\IM{\mathbb{M}}

\def\IP{\mathbb{P}}
\def\IQ{\mathbb{Q}}
\def\IR{{\mathbb{R}}}

\def\IZ{{\mathbb{Z}}}

\def\fo{\mathfrak{o}}

\def\fs{\mathfrak{s}}

\def\fs{\mathfrak{s}}

\def\fu{\mathfrak{u}}

\def\fF{\mathfrak{F}}
\def\fL{\mathfrak{L}}

\def\fX{\mathfrak{X}}
\def\fY{\mathfrak{Y}}
\def\fX{\mathfrak{X}}

\def\rmk#1{\bigskip\noindent{\bf Remarks} }

\title{Conway Subgroup Symmetric Compactifications Of Heterotic String}

\author{Jeffrey A. Harvey$^1$ and Gregory W.~Moore$^2$ \\
$^1$ Enrico Fermi Institute and Department of Physics\\
$~~ $University of Chicago \\
$~~ $5620 Ellis Ave., Chicago IL 60637 \\
$^2$ NHETC and
$~~$Department of Physics and Astronomy, Rutgers University \\
$~~$126 Frelinghuysen Rd., Piscataway NJ 08855, USA\\
\\
{\tt j-harvey@uchicago.edu, gmoore@physics.rutgers.edu } }

\abstract{We investigate special
compactifications of the heterotic string for which the space of  half-BPS states
is, in a natural way, a representation of various subgroups of the Conway
group.  These compactifications provide a useful framework for analyzing the action
of some of the large symmetry groups appearing in discussions of Moonshine in the physics
literature.  We investigate toroidal compactifications of heterotic string with sixteen
supersymmetries as well as asymmetric toroidal orbifolds with
 $N=2$ supersymmetry in four dimensions that arise as $K3 \times T^2$ compactifications.
The latter Conway subgroup symmetric compactifications of the heterotic string might
have some interesting implications for D-brane bound states on Calabi-Yau manifolds.\vskip 0.1in
\today}


\begin{document}

\section{Introduction And Conclusion}\label{sec:Intro}

This paper discusses discrete symmetries of some special toroidal, and toroidal orbifold,
compactifications of the heterotic string.  The construction of the toroidal compactifications is
described in section \ref{subsec:LatticeLemma}. The construction is a simple application of results of  Nikulin \cite{Nikulin}.
There are three motivations for discussing this topic.

The first motivation
concerns Moonshine, old and new. As is very well known, the
original Moonshine conjectures of Conway and Norton \cite{conwaynorton},
associated with the Monster group and the modular $j$ function,
led to some very interesting developments
in string theory and in mathematics including the construction of an explicit Vertex Operator Algebra (VOA) or holomorphic
Conformal Field Theory (CFT) with Monster symmetry in \cite{flm} and the proof of the genus zero property of Monstrous Moonshine by Borcherds \cite{bormoon}.  In recent years various new types of Moonshine conjectures
have again caught the attention
of string theorists and mathematicians \cite{UM,UMNL,onan,Eguchi:2010ej,Harvey:2015mca}.
Some of these conjectures have now been proved \cite{DGO,Gannon,griffin}, but the proofs are not constructive and the full implications
of these new Moonshine results both for string theory and for mathematics remain to be understood. In particular, it remains to be seen whether
or not there are VOA and/or CFT constructions which underlie these new examples of Moonshine.
The present paper studies CFTs associated with heterotic string theory compactifications with large discrete symmetries and is thus potentially relevant to the new Moonshine phenomena.

A second motivation for this paper is the evidence for Moonshine associated to the Mathieu group $M_{24}$ that was recently observed in the computation of refined
DT-type invariants for K3 surfaces \cite{Katz:2014uaa}. The relevance of the present paper to the considerations of
\cite{Katz:2014uaa}  is provided
courtesy of heterotic-type II duality. (For a review, see \cite{Aspinwall:1996mn}.)
Viewed through the lens of heterotic-type II duality, the invariants
of \cite{Katz:2014uaa}  are counting perturbative BPS states in the heterotic
dual so it is natural to ask whether there are heterotic compactifications in which the perturbative BPS states form a representation for
$M_{24}$ or closely related groups.
In Appendix \ref{App:NoMoonshine} of this paper we will give
some  strong evidence that, regrettably, there is in fact no natural $M_{24}$ representation underlying the
DT invariants computed in \cite{Katz:2014uaa}.
The $O(4)$ representations studied in \cite{Katz:2014uaa} are organized by $p^2$, rather than $p$ (where $p$ is the
Narain lattice vector).
For this reason they are not sensitive to the specific crystal structure of the Narain lattice and the degeneracies
are  more simply understood in terms of the natural $O(20) \times O(4)$ symmetry of dimensional reduction to six
noncompact dimensions: The $O(4)$ is the massless little group in six dimensions while the $O(20)$ rotates all the
``internal'' left-moving bosonic fields of the heterotic string.

In spite of this negative, and disappointing, conclusion
 our considerations do raise the interesting issue of the relation of symmetries present in a worldsheet,
or perturbative, analysis of a string theory  to   symmetries of nonperturbative
states.   For example, it was shown in \cite{Gaberdiel:2011fg}  that any automorphism of a
(smooth) K3 sigma model
%
%
%
is a subgroup of the Conway group stabilizing
a four-plane in the Leech lattice.  It is interesting to ask whether this classification, based on the worldsheet conformal
structure, also extends to symmetries of the full string theory including nonperturbative states. The study of heterotic string on $T^4$
provides a tool to start addressing this question since perturbative BPS states in the heterotic string
(first studied in  \cite{Dabholkar:1989jt}) map to nonperturbative states
of the type II string on $K3$ under string duality. Indeed, heterotic/type II duality implies that in this case the
worldsheet symmetries identified in \cite{Gaberdiel:2011fg} do indeed extend to the nonperturbative BPS sector.
In fact, the relevant mathematical version of this statement has been rigorously proven in \cite{Huybrechts:2013iwa}.

The third and final motivation for this paper is based on a very natural approach to giving a conceptual
framework for the Mathieu Moonshine of \cite{Eguchi:2010ej}. The basic idea
is to  identify $M_{24}$ as an automorphism group (or a distinguished subgroup thereof)
of the algebra of BPS states discussed in \cite{Harvey:1995fq,Harvey:1996gc}.
In the present context there are actually several ways in which this general idea could
be implemented. One might try to consider the algebra of BPS states of the
type IIA string compactified to six Minkowski
dimensions along a K3 surface. This would involve considerations of
wrapped D4-D2-D0 brane bound states (or objects in the derived category of K3) mentioned above.
Again using the lens of heterotic-type II duality, one  might try to consider instead
the algebra of perturbative BPS states
of the toroidally compactified heterotic dual theory. The latter has the advantage that the algebra
then has a concrete and computable definition in terms of vertex operator algebras
\cite{Harvey:1996gc}.  (The ``correspondence conjecture''
of \cite{Harvey:1996gc} asserts that the algebra constructed using vertex operators has a geometrical construction on the type II side, the latter
being inspired by work of Nakajima.)
This paper focuses on the heterotic string. Unsurprisingly,
in view of the work of \cite{Gaberdiel:2011fg}, we will show that it is possible to
formulate heterotic toroidal CFTs with discrete symmetries related to subgroups of
the Conway group. As stressed by \cite{Gaberdiel:2011fg} the relevant list of symmetries
is  not naturally related to $M_{24}$. That leaves the logical possibility
 that there are
extra symmetries of the BPS sector, not present in the full CFT. Such ``extra symmetries''
might well exist and we have not ruled them out. Their existence would be extremely interesting.
%
%
However, we can say that the  the considerations of  Appendix \ref{App:NoMoonshine}
indicate that $M_{24}$ cannot be such a group of ``extra symmetries,''  even when restricting
attention to the BPS sector.

From the viewpoint of explaining Mathieu Moonshine our results are thus largely negative.
We hasten to add that our considerations do not
 exclude the possible relevance of BPS algebras to the explanation of Mathieu
Moonshine. Rather, they narrow the search
for situations in which such algebras could be relevant. For example, one natural and
unexplored direction is to consider perturbative type IIA BPS states in a compactification
on $\IR^{1,4}\times K3 \times S^1$ or (what seems to us a more likely scenario) on
$\IR^{1,1}/II^{1,1} \times K3 \times \fX$ for a suitable compact space $\fX$. It seems to us
that this is a potentially interesting line of future research.

Given the toroidal compactifications with large discrete symmetries related to the Conway
group we can still do some interesting things with them, again in light of heterotic/type II
duality. Namely, we can consider orbifolds of the heterotic string to $T^2 \times K3$
compactifications that still preserve large subgroups of the Conway group. We then expect
these to imply interesting generalizations of the automorphisms of the derived category
of K3 explored in \cite{Huybrechts:2013iwa}. Motivated by this we have constructed
several such (asymmetric) orbifolds. As it turns out, this is no easy task. The methods
and results are outlined in section \ref{sec:ExampleCSS} and some of the many details are given
in Appendices \ref{app:UntwistedSector},\ref{App:GolaySteiner},\ref{App:Details-222},\ref{App:Details-224}.
The key properties of the models are summarized in Tables \ref{table:HM222} and
\ref{table:HM224} below.
We suggest some potential implications of our results for Calabi-Yau geometry and directions for
further research in Section \ref{sec:HetTypeII}.

\section*{Acknowledgements}

We thank T. Banks, T. Bridgeland, M. Cheng, E. Diaconescu, R. Donagi,
 G. H\"ohn, G. Mason, D. Morrison, K. Narain, N. Paquette, D. Park, B. Pioline, M. Rocek,
 N. Seiberg,  and W. Taylor
for very useful discussions and correspondence.  JH and GM gratefully acknowledge
the hospitality of the Aspen Center for Physics  (under
NSF Grant No. PHY-1066293) where this work was initiated.
They also thank
the Perimeter Institute for Theoretical Physics and the
Institute For Advanced Study for hospitality.  We have made use of the GAP package for finite group theory in our analysis \cite{GAP} as well as the Magma Computer Algebra system which was
made available through a grant from the Simons Foundation. JH acknowledges support from the NSF \footnote{Any opinions, findings, and conclusions or recommendations expressed in this material are those of the author(s) and do not necessarily reflect the views of the National Science Foundation.} under grant PHY 1520748 and from the Simons Foundation (\#399639).
GM is supported by the DOE under grant
DOE-SC0010008 to Rutgers.

\section{Conway Subgroup Symmetric Compactifications}\label{sec:CSS-Compact}

\subsection{Recollections On Narain Compactification}\label{subsec:NarainReview}

We consider toroidal compactification of the heterotic
string on
\be\label{eq:Spacetime-d}
\IM^{1,1+d} \times T^{8-d}.
\ee
 As is well-known,
the lattice of zero modes of the worldsheet coordinates
forms an even unimodular lattice of signature $(24-d;8-d)$
embedded in a fixed pseudo-Euclidean space of the same signature.
We denote the lattice as
\be
\Gamma^{24-d;8-d} \subset \IR^{24-d;8-d},
\ee
or sometimes just by $\Gamma$ when $d$ is understood. The projections of a vector $p\in \Gamma^{24-d;8-d}$
onto the definite subspaces $\IR^{24-d;0}$
and $\IR^{0;8-d}$, will be denoted by $p_L$ and $p_R$, respectively.
\footnote{We will generically denote vectors in $\IR^{24-d;8-d}$
by $(x;y)$ where the projection to the negative definite
subspace $\IR^{24-d;0}$ is $x$ and
the projection to the positive definite subspace $\IR^{0;8-d}$ is $y$.
If $L$ is a positive definite lattice embedded in Euclidean
space $\IR^{24-d}$ we denote by $(L;0)$ the sublattice of
$\IR^{24-d;8-d}$ consisting of vectors  $(\ell;0)$ with $\ell\in L$.
Thus, as an abstract lattice $(L;0)$ is isomorphic to $L(-1)$. We let $II^{p,q}$ denote
the even unimodular lattice of signature $(p,q)$, so $p-q=0~\mod~ 8$ and we take $p,q>0$. }
Orthogonal rotations of these definite subspaces are
symmetries of the worldsheet conformal field theory and
hence the moduli space of toroidal theories is the
Narain space
\be
O_{\IZ}(II^{24-d;8-d}) \backslash O_{\IR}(24-d;8-d)/(O_{\IR}(24-d) \times O_{\IR}(8-d)).
\ee
%
%

At points in Narain moduli space where one of the groups
\be
G_L  = {\rm Aut}(\Gamma^{24-d;8-d})\cap O_{\IR}(24-d) \qquad \qquad
G_R  = {\rm Aut}(\Gamma^{24-d;8-d})\cap O_{\IR}(8-d)
\ee
is nontrivial there are orbifold singularities, and there is a lift of
crystal symmetry $G_L \times G_R$ to groups $\tilde {G_L} \times \tilde {G_R}$  which are global symmetries of the worldsheet
theory \cite{Harvey:2017rko}.
%
%
A famous example of such discrete symmetries
are the Weyl-group symmetries at subloci of Narain moduli space at which
there are enhanced non-abelian gauge symmetry. We are interested here
in other kinds of crystal symmetries that are not of this type. We
call them ``Conway subgroup symmetries.''

The simplest example of a Conway subgroup symmetry arises for $d=0$,
that is, toroidal compactification of the heterotic string to two
spacetime dimensions. A distinguished point in Narain moduli space
corresponds to the embedded even unimodular lattice:
\footnote{In what follows  we will use a concrete model of the $E_8$ lattice $\Gamma_8 \subset \IR^8$
as the set of vectors $(y_1,\dots, y_8)$ where the coordinates are either all
integral or all integral plus $1/2$ and in both cases the sum of the coordinates is even.
We will also use a similar description of the Leech lattice $\Lambda$.
See Appendix \ref{App:GolaySteiner} for some relevant definitions.}
\be\label{eq:SpecialPoint}
\Gamma^{24;8}_* := (\Lambda;0 ) \oplus (0; \Gamma_8).
\ee
The crystal symmetry for this compactification is $\Co0 \times W(E_8)$,
where $\Co0$ is the Conway group, by definition the group of automorphisms
of the Leech lattice $\Lambda$ \cite{Conway1}.

Most of what follows in this paper applies equally well to the other 23 distinguished points obtained by replacing the Leech lattice by one of the Niemeier lattices. Indeed this democracy among the 24 Niemeier lattices played an important role in \cite{Cheng:2016org,Kachru:2016ttg}.

\subsection{A Lattice Lemma}\label{subsec:LatticeLemma}

Our first goal is to construct lattices $\Gamma^{24-d;8-d}$ which do not have enhanced gauge symmetry, that is have no
points $(p_L,0) \in \Gamma^{24-d;8-d}$ with $p_L^2=2$ and yet have enhanced discrete symmetries. To do so we need
need the following result in the theory of lattices:

\bigskip
\noindent
\textbf{Lemma}: Suppose we have two primitively embedded sub-lattices $\fF_R\subset \Gamma_8$ and $\fF_L \subset \Lambda$
where $\Gamma_8$ is the $E_8$ lattice and $\Lambda$ is the Leech lattice,
and suppose that $\fF_R $ and $\fF_L $ are of rank $d$ and  isometric.   Then we can
construct an even unimodular lattice $\Gamma \cong II^{24-d;8-d}$
and an embedding of $\Gamma $ into $\IR^{24-d;8-d}$   such that
\be
{\rm Fix}(\fF_L) \times {\rm Fix}(\fF_R)    \subset {\rm Co}_0 \times  W(E_8)
\ee
is a crystallographic symmetry of $\Gamma$ with ${\rm Fix}(\fF_L)\subset O(24-d)$
and ${\rm Fix}(\fF_R) \subset O(8-d)$.

\bigskip
\noindent
\emph{Proof}:  Let us begin by recalling a standard fact from lattice theory.
(See \cite{MirandaMorrison,Nikulin} for further explanation.) If $\fL$ is
an even integral lattice then we can define the discriminant group $\CD_\pm(\fL)$
as follows. We consider the dual lattice $\fL^\vee \subset \fL\otimes \IQ $
and the finite abelian group $\CD(\fL):=\fL^\vee/\fL$. This finite group  inherits a pair of
quadratic functions $q_\pm: \CD(\fL) \rightarrow \IQ/2\IZ $ using the inner product on $\fL^\vee$:
\be
q_\pm(\bar v):=  \pm v^2 ~ \mod ~2\IZ
\ee
where $\bar v$ has a representative $v\in \fL^\vee$. That is $\bar v = [v]$.
Note that $q_\pm(\bar v)$ does not depend on the representative $v$.
The phrase ``quadratic function'' means that
\be
q_\pm(\bar v_1 + \bar v_2) - q_\pm(\bar v_1) - q_\pm(\bar v_2) + q_\pm(0) := 2 b_\pm(\bar v_1, \bar v_2)
\ee
defines a bilinear form $b_\pm$ valued in $\IQ/\IZ$. We denote by $\CD_\pm(\fL)$  the finite group
equipped with the quadratic function $q_\pm$. Sometimes we will write $q_{\CD_\pm(\fL)}$ when we want to emphasize the lattice.

Now suppose that  $\Gamma$ is an even
unimodular lattice and  $\fF \subset \Gamma$ is a
primitively embedded sublattice.  Primitively embedded means that the Abelian group $\Gamma/\fF$ is free, that is it has no
elements of finite order. Then in \cite{MirandaMorrison,Nikulin}  it is shown that there is a canonical
isometric isomorphism
\be
\psi: \CD_+(\fF) \rightarrow \CD_-(\fF^\perp)
\ee
where ``isometric'' means $q_-(\psi(\bar v)) = q_+(\bar v)$.

In particular, given a primitively embedded sublattice
 $\fF_L \subset \Lambda$   there is an isomorphism of discriminant groups
\be
\psi_L: \CD_+(\fF_L^\perp) \to \CD_-(\fF_L)
\ee
where $\fF_L^\perp \subset \Lambda$ is the orthogonal complement of $\fF_L$ within $\Lambda$.
Similarly, there is an isomorphism
\be
\psi_R: \CD_-(\fF_R) \to \CD_+(\fF_R^\perp) \, .
\ee

Now we define a quadratic space $V$ by
\be
V =  \Lambda\otimes \IQ  \oplus \Gamma_8 \otimes \IQ \cong \IQ^{24;8}
\ee
writing vectors as $(x;y)$ with $x\in  \Lambda\otimes \IQ $ and
$y\in \Gamma_8 \otimes \IQ $ with
\be
(x;y)^2 = - x^2 + y^2
\ee
We can of course further extend scalars from $\IQ$ to $\IR$.

Similarly define the subspaces of $V$:
\be
W_d :=   \fF_L^\perp \otimes \IQ \oplus   \fF_R^\perp \otimes \IQ \subset V
\ee
$W_d$ is a quadratic space isomorphic to $\IQ^{24-d;8-d}$.
\footnote{Note that which subspace of $V$ we get depends on $\fF_L$ and
$\fF_R$, and hence is less canonical than $V$ itself.}

Now we choose an isometry  between $\fF_L$ and $\fF_R$.
This will induce an isomorphism
\be
 \psi_{RL}: \CD_-(\fF_L) \to \CD_-(\fF_R)
\ee
allowing us to define an isomorphism
\be
\psi: \CD_+(\fF_L^\perp) \to \CD_+(\fF_R^\perp)
\ee
by
\be
\psi = \psi_R \circ \psi_{RL} \circ \psi_L.
\ee

Next define an embedded lattice in $W_d$:
\be
\tilde \Gamma := \left( \fF_L^\perp\right)^\vee \oplus \left( \fF_R^\perp\right)^\vee\subset W_d
\ee
and also define a sublattice of $\tilde \Gamma$:
\be
\Gamma \subset \tilde \Gamma
\ee
to be the set of vectors $(x;y) \in \tilde\Gamma$ such that $\psi(\bar x) = \bar y$.

It is a standard result that the lattice $\Gamma$ constructed this way is even unimodular
\cite{MirandaMorrison,Nikulin}, but for completeness let us recall the proof.
The fact that it is an integral even lattice is easy to demonstrate: If $(x;y) \in \Gamma$ then
\be
\begin{split}
(x;y)^2~ \mod ~2 & = -x^2 + y^2~ \mod~ 2 \\
& = -q_{\CD_+(\fF_L^\perp)}(\bar x) + q_{\CD_+(\fF_R^\perp)}(\bar y) \\
& = 0  \\
\end{split}
\ee
Since the bilinear form can be derived from the quadratic function it follows that the inner product
of any two vectors in $\Gamma$ is integral. Similarly, we can show it is unimodular
as follows: Suppose $(u;v) \in \Gamma^\vee$. Then for all $(s;t) \in \Gamma$ we have
$-u\cdot s + v \cdot t =0 ~\mod ~1$. So, taking $t=0$ and then $s=0$ shows that
$(u; v ) \in  \tilde\Gamma$. we have
\be
-\bar u \cdot \bar s + \bar v \cdot \bar t = 0 ~ \mod ~ 1
\ee
and hence
\be
(\bar u - \psi^{-1}(\bar v) ) \cdot \bar s = 0 ~ \mod ~ 1.
\ee
Now, since $\CD_+(\fF_L^\perp)$ is a \underline{non-degenerate} quadratic space
we must have $\bar u - \psi^{-1}(\bar v) = 0 $ and hence
$\psi(\bar u) = \bar v$ and hence $(u,v) \in \Gamma$.

Choosing an isomorphism $W_d \cong \IQ^{24-d;8-d}$
and extending scalars to $\IR$ we have constructed an embedding of $II^{24-d;8-d}$ into $\IR^{24-d;8-d}$.

Now, let $G_L={\rm Fix}(\fF_L) \subset {\rm Aut}(\Lambda) = {\rm Co}_0$ and similarly $G_R$.
Extending scalars we can embed $G_L \subset O(\Lambda\otimes \IQ) \subset O(\Lambda\otimes \IR) \cong O(24)$,
and similarly $G_R$. Now we claim that if $x \in (\fF_L^\perp)^\vee\subset \Lambda\otimes\IQ$ and $g\in G_L$
then
\be\label{eq:Prsv}
g \cdot x -x \in \fF_L^\perp
\ee
(\emph{a priori} we only know that $g \cdot x -x \in (\fF_L^\perp)^\vee$).
This follows since for any $x\in (\fF_L^\perp)^\vee$ there is an $x' \in \fF_L^\vee$
so that $x \oplus x' \in \Lambda$. But then
\be
g\cdot x - x = g\cdot (x \oplus x') - (x\oplus x') \in \Lambda
\ee
But now \eqref{eq:Prsv} implies that   in the discriminant group $g(\bar x) = \bar x$ for any $x \in (\fF_L^\perp)^\vee$ and $g\in G_L$.
Entirely similar remarks apply to $g\in G_R$ and $y\in (\fF_R^\perp)^\vee$.

Now we define the action of $G_L \times G_R$ on $\tilde \Gamma$ by
\be
(g_L,g_R) \cdot (x;y) := (g_L \cdot x; g_R \cdot y)
\ee
Moreover, as we have seen $g_L \bar x = \bar x$ and $g_R \bar y = \bar y$.
Therefore if $(x;y) \in \Gamma$ then
\be
\psi( g_L \bar x) = \psi(\bar x) = \psi(\bar y) = \psi( g_R \bar y)
\ee
and hence the action of $G_L \times G_R$ preserves $\Gamma$. This concludes the proof $\spadesuit$

In essence the lattice $\Gamma$ is simply given by adding suitable glue vectors  to $\fF_L^\perp \oplus  \fF_R^\perp + \cdots $.
Given the above Lemma we now define the \emph{Conway subgroup symmetric compactifications
of the heterotic string} to be defined by those points in Narain moduli space
associated with such pairs of isometric sublattices $\fF_L \subset \Lambda$ and
$\fF_R \subset \Gamma_8$. We refer to these as CSS compactifications for short.

\bigskip
\noindent
\textbf{Remarks}:

\begin{enumerate}

\item H\"ohn and Mason   have tabulated the $290$ isomorphism classes of
sublattices $\fL_L \subset \Lambda $ such that the subgroup ${\rm Fix}(\fF_L) \subset \Co0$
fixing all vectors of $\fF_L$ is nontrivial \cite{HM}. The reader
of this paper who wants to follow the details of the analysis will find it useful to have Table 1 and supplementary Table 2 of \cite{HM} handy.
Said reader might also note the versions two and three of \cite{HM} are useful for different purposes.
We will refer to entries with number $\#$ in their table as HM$\#$ for short.

\item For $d=4$ there is a kind of converse of this result
due to Gaberdiel, Hohenegger, and Volpato \cite{Gaberdiel:2011fg}.
They have shown that $G\subset O_{\IZ}(II^{20,4})$ fixes a positive
definite four-dimensional subspace of $\IR^{20,4} = II^{20,4}\otimes \IR$
iff  $G$ is a subgroup of $\Co0$ fixing a sublattice of $\Lambda$ of
rank at least four.
Although we will not use this result directly, it was important to our thinking.
\bigskip
\bigskip

\end{enumerate}

\section{Examples Of CSS Compactifications}\label{sec:ExampleCSS}

We now turn to compactifications of the heterotic string on
$T^2 \times K3$ preserving four-dimensional $N=2$
supersymmetry. These are notable for having
type IIA
dual compactifications on $K3$-fibered Calabi-Yau threefolds \footnote{We are not aware of any general proof that all heterotic compactifications
with $N=2$ supersymmetry in four dimensions have type II duals when one includes the kind of non-geometrical compactifications used here involving
asymmetric orbifold constructions}.
(For a review, see \cite{Aspinwall:1996mn}.) We would like
to consider such compactifications with Conway subgroup symmetry.

\subsection{Right-moving Lattice $\fF_R$ And Orbifold Action On Right-Movers}

The natural way to produce CSS  compactifications
of the heterotic string on $T^2 \times K3$ is to consider suitable orbifolds
of CSS compactifications on $T^6$. Compactification on such tori is the case $d=2$ in
\eqref{eq:Spacetime-d}. Therefore we can produce models by
considering the rank two sublattices
$\fF_L \subset \Lambda$ in Tables 1 and 2 of \cite{HM}
and then searching for isometric primitively embedded sublattices $\fF_R \subset \Gamma_8$
such that there is a subgroup $G_R \subset  {\rm Fix}(\fF_R) \subset W(E_8)$
so that if $T^6 = \IR^6/\fF_R^\perp$ then
\be
T^6/G_R  \cong T^2 \times S
\ee
with $S$ an orbifold limit of a K3 surface.

For simplicity we will limit attention here to the simplest
case of $G_R \cong \IZ_2$. We therefore seek an involution
in the Weyl group of $E_8$. Moreover, our ``K3 surface'' will be the
orbifold $T^4/\IZ_2$ and hence we need an involution with precisely four $-1$
eigenvalues, that is, one such that the character in the natural
eight-dimensional representation is zero.
There are two conjugacy classes of such involutions. We can describe
one representative of the first conjugacy class as a sign flip on the first four coordinates
(using the standard model for $\Gamma_8$ mentioned near \eqref{eq:SpecialPoint}):
\be
\sigma_1 =  (-1^4,+1^4) \, .
\ee
A representative of the second conjugacy class acts on the coordinates as a matrix:
\be
\sigma_2 = \begin{pmatrix}  H & 0  \\  0 & H \\   \end{pmatrix}
\ee
where $H$ is the famous Hadamard matrix
\be
H := \half \begin{pmatrix}
1 & 1 & 1 & 1 \\
1 & -1 & 1 & -1 \\
1 & 1 & -1 & -1 \\
1 & -1 & -1 & 1 \\
\end{pmatrix} \, .
\ee
It will be useful later to note that the four vectors  $u_i \in \Gamma_8$ given by
\be
\begin{split} \label{righthada}
u_1 &= (1/2)(1,1,1,-1,1,1,1,-1) \\
u_2 &= (1/2)(1,1,1,-1,-1,-1,-1,1) \\
u_3 &= (1,0,0,1,0,0,0,0) \\
u_4 &= (0,0,0,0,1,0,0,1)
\end{split}
\ee
form a basis for the rank $4$ sublattice of $\Gamma_8$ which is fixed by the Hadamard involution.

It is now a simple, albeit tedious matter to enumerate the possible
rank two sublattices $\fF_R\subset \Gamma_8$ fixed by the above two involutions
and compare with the sublattices $\fF_L$ in the tables of \cite{HM}.
Of the $51$ rank two HM classes we find all but $6$. The missing
classes are $\# 223, \#227, \# 232, \# 237, \# 240, \# 246$. with
Gram matrices:
\be
\begin{pmatrix} 4 & 1 \\ 1 & 4 \\ \end{pmatrix},
\begin{pmatrix} 4 & 1 \\ 1 & 6 \\ \end{pmatrix},
\begin{pmatrix} 4 & 2 \\ 2 & 8 \\ \end{pmatrix},
\begin{pmatrix} 4 & 2 \\ 2 & 16 \\ \end{pmatrix},
\begin{pmatrix} 6 & 3 \\ 3 & 12 \\ \end{pmatrix},
\begin{pmatrix} 8 & 2 \\ 2 & 8 \\ \end{pmatrix},
\ee
respectively.  It is possible that these missing classes can
be captured by considering other orbifold limits of K3 surfaces,
but we have not investigated this possibility.

We also need to construct
a \underline{primitive} embedding of a sublattice $\fF_R$ into  $\Gamma_8$
where $\fF_R$ has the desired Gram matrix. The existence of such an embedding is
guaranteed by Theorem 1.12.2 of \cite{Nikulin} and in explicit examples discussed below
we will construct these primitive embeddings explicitly.

However there is no guarantee that such a primitive embedding can be chosen to lie within the invariant subspace of one
of the involutions $\sigma_1, \sigma_2$.
For the Gram matrix $Q^{224}$ in the following subsection  there are obvious choices of orthogonal square-length four
vectors fixed by the involution $\sigma_1$, but one can check that no such pair generates
a primitive sublattice of $\Gamma_8$. In fact for both of the examples in the following subsection we can find a primitive embedding and choose basis vectors for $\fF_R$ that are
in the $\sigma_2$-invariant subspace.

\subsection{Left-moving Lattice $\fF_L$ }

We now focus on two illustrative examples chosen to have large CSS groups as stabilizers. The Gram
matrices are \#222 and \#224 in \cite{HM} and are given by
\be\label{eq:ourgrams}
Q^{222}=\begin{pmatrix} 4 & -2 \\ -2 & 4 \end{pmatrix} \qquad  Q^{224} = \begin{pmatrix} 4 & 0 \\ 0 & 4 \end{pmatrix} \, .
\ee
According to Table 1 of \cite{HM} the group $\Fix(\fF_L)$ with Gram matrix  $Q^{222}$ is a subgroup of ${\rm Co}_0$
 isomorphic to $U_6(2)$. For definiteness, we will choose $\fF_L$ to be the sublattice of the Leech lattice generated by
 the vectors  $v_1$ and $-v_3$ of \eqref{eq:LeechBasis}.

Similarly, for $\fF_L$ with Gram matrix $Q^{224}$ Table 1 of \cite{HM} says the stabilizer group
 ${\rm Fix}(\fF_L) \cong 2^{10}.M_{22}$, and
moreover it is a subgroup of the monomial subgroup $2^{12}:M_{24}$
of the Leech lattice, described in Appendix \ref{App:GolaySteiner}.
As described in Appendix \ref{App:GolaySteiner} we are
using a presentation of the Leech lattice so that
$\{1,\dots, 8\}$ is a set in the Steiner system $S(5,8,24)$.
For definiteness, we choose $\fF_L$ to be the sublattice generated by
the vectors $v_1$ and $v_2$ of \eqref{eq:LeechBasis}.

See Appendices  \ref{App:Details-222} and \ref{App:Details-224}  for the detailed construction
of the entire Narain lattice $\Gamma$ of signature $(22,6)$  for each of these two choices of $\fF_L$.

\subsection{Orbifold Action On Left-Movers}

Once one has constructed $\Gamma$ and chosen the action of the involution on the
right-moving coordinates it remains to choose the involution
on the left-moving coordinates. This will be of the form
\be
x \to g_L x + \delta
\ee
where $g_L \in {\rm Fix}(\fF_L)$ is an involution and $\delta$ is in $\Lambda \otimes \IQ$
such that $g_L \delta + \delta \in \Lambda$. There
are four conjugacy classes of involutions in $\Co0$ with characters
$-24,8,-8,0$ in the $24$-dimensional irreducible representation. Thus in a diagonal
basis $g_L$ will have $24,8,16,12$ eigenvalues $-1$, respectively.
Since we must have at least $4$ eigenvalues $+1$, to produce
a compactification on $T^2 \times K3$, (these account for the $T^2$ and the $\IR^2$ directions)
the involution $g_L$ must be in one of the conjugacy classes with $8$, $16$, or $12$ eigenvalues equal
to $-1$.

We now define three concrete models corresponding to three choices of $(g_L, \delta)$. See Appendix \ref{App:GolaySteiner}
for the definitions of $\CC$-set, octad and dodecad.

\begin{enumerate}

\item $A$:  $g_L$ is a sign-flip on a $\CC$-set fixing $\fF_L$. For both $Q^{222}$ and $Q^{224}$
we can choose an octad so that that sign flip fixes $x_1, x_2, x_3$. For example the flips
associated with $S_{22}$ or $S_{23}$ will do.
Moreover we take the shift vector to be $\delta =0$. Note that there is a variant of this model
with a shift vector $\delta = \half v_4$ with $v_4 \in \Lambda$ with $v_4^2=4$, but we will analyze only the model
with vanishing shift vector.


\item $B$: $g_L$ is a sign-flip on an octad-complement of a $\CC$-set that leaves $\fF_L$ pointwise fixed.
For example, a sign flip on the complement of $S_8$ will fix coordinates $x_1, x_2, x_3$ and hence will
fix $\fF_L$ for both cases $Q^{222}$ and $Q^{224}$.

\item $C$: $g_L$ is a sign-flip on a dodecad $\CC$-set that leaves $x_1,x_2$ invariant.
To be concrete for $Q^{224}$ we choose the dodecad in \eqref{eq:Dodecad} so:
\be
\begin{split}
g_L^C x_i & = x_i \qquad  i \in \{ 1,2,5,7,10,12,13,14,17,18,22,24 \} \\
g_L^C x_i & = - x_i \qquad  i \in \{ 3, 4, 6, 8, 9, 11, 15, 16, 19, 20, 21, 23\}\\
\end{split}
\ee
and for $Q^{222}$ we choose the dodecad in \eqref{eq:Dodecad}
\be
\begin{split}
g_L^C x_i & = x_i \qquad  i \in \{ 1,2,3,4,9,10,15,16,19,20,23,24 \} \\
g_L^C x_i & = - x_i \qquad  i \in \{ 5,6,7,8,11,12,13,14,17,18,21,22\}\\
\end{split}
\ee
Level matching requires that we choose $\delta = \half v$ with $v^2=2~\mod~ 4$.
In order to preserve a large crystallographic symmetry acting on the left we choose
$v$ to be in the right-moving part of the lattice and invariant under the Hadamard involution.
Thus we can choose $v$ to be any of the $u_i$ in \eqref{righthada}.

%
%

\end{enumerate}

The orbifold models $A, B, C$ all satisfy standard level-matching constraints required for modular invariance.
In addition  one should address the
``DTF'' criterion of \cite{Harvey:2017rko} (which extends earlier treatments in \cite{LepowskyCalculus,Narain:1986qm,NSVII}.)
Namely if there is a vector  $p\in \Gamma$ such that
\be\label{eq:NSV-Mod2}
( p , g \cdot p ) = 1~ \mod ~2 \, .
\ee
then, as described under equation $(2.17)$ of \cite{Harvey:2017rko} either $\hat g \vert p \rangle \not= \vert g\cdot p \rangle$
for some $p\in \Gamma^g$ or the lifted automorphism $\hat g$ on the CFT
will be order four. When the DTF criterion \eqref{eq:NSV-Mod2} holds we will use the canonical lift of equations $(6.35)$ and $(6.36)$ of
\cite{Harvey:2017rko}.
After some work one can show that for the involutions of type $A$ and $B$ the DTF condition \eqref{eq:NSV-Mod2} does
indeed hold for involutions of type A,B for both $Q^{222}$ and $Q^{224}$.  As a result models A,B are actually $\IZ_4$ orbifolds.
Strangely, for $C$ the condition \eqref{eq:NSV-Mod2} again holds for $Q^{222}$ but not for $Q^{224}$, so only in
the last case do we have a  true $\IZ_2$ orbifold.
Altogether we will analyze six $K3 \times T^2$ orbifolds of CSS compactifications labelled by
the two choices of Gram matrix ($\#222$ and $\#224$) and three choices of left-moving lattice involution ($A,B,C$).

\subsection{Massless States In The Untwisted Sector}

It is straightforward  to compute the untwisted sector massless spectrum of these models.
(For some details see Appendix \ref{app:UntwistedSector}.)   Let $n_-=8,16,12$ be the number of $-1$
eigenvalues of $g_L$ in cases $A,B,C$, respectively. In the untwisted
sector one finds one  $N=2$ supergravity multiplet,  $23-n_-$ $U(1)$
vectormultiplets, and $n_-$ hypermultiplets. The massless scalars
are moduli.  Before dividing by duality
symmetries we have a vector-multiplet moduli space
\be\label{eq:VM-modulispace}
\CM_{vm} = \frac{SL(2,\IR)}{SO(2)} \times \frac{O(2,22-n_-)}{O(2) \times O(22-n_-)}
\ee
and a hypermultiplet moduli space
\be
\CM_{hm} =  \frac{O(4,n_-)}{O(4) \times O(n_-)}
\ee

The computation of the spectrum in the twisted sector for model C is
also straightforward for the simple reason that
the ground state energy in the twisted sector is positive
so there are no massless twisted sector states and hence the full massless
spectrum consists of $11$ vectormultiplets and $12$ hypermultiplets.

\subsection{Massless States In The Twisted Sectors}

For models $A,B$ with order four lifts there are massless states in the
twisted sectors and we need to work much harder. To compute the
massless twisted sector we begin with the trace in the unprojected
theory:   $\Tr_{\CH} \hat g q^{L_0 - c/24}
\bar q^{\tilde L_0 - \tilde c/24}$ where for the canonical lift  we have
\be
\langle p \vert \hat g \vert p \rangle = \begin{cases} 1 & p \in \Gamma^g \\  0 &  {\rm else} \\
\end{cases}
\ee
Now let $Z(\hat g^x, \epsilon_1; \hat g^y , \epsilon_2) $ be the twisted partition
function with spin structure specified by $(\epsilon_1, \epsilon_2)$ (relative
to the canonical RR spin structure)  and twist $\hat g^y$ in the space and
$\hat g^x$ in the time direction. Including orbifold and GSO projection the
$\hat g$-twisted NS sector partition function is:
\be
Z(\CH_{\hat g}) = \frac{1}{2} \cdot \frac{1}{4}
\Biggl[  \sum_{x=0}^3 \sum_{\epsilon= \pm } Z( \hat g^x, \epsilon; \hat g, -) \Biggr].
\ee

Now, let the number of $-$ eigenvalues be $n_-$ on left-movers and $\tilde n_-$ on
right-movers. For our example $\tilde n_- = 4$ but we will leave the formulae
general for a little while, merely assuming $\tilde n_- >0$. Then the spatial
twist $\hat g, -$ means that there are $\tilde n_-$ real fermions with periodic
boundary conditions. Therefore $Z(\hat g, -; \hat g, - )$ and all its images
under $\tau \to \tau +1$ vanish. This leaves four terms:
\be
Z(\CH_{\hat g}) = \frac{1}{2} \cdot \frac{1}{4}
\Biggl[  Z(1,-; \hat g, - )  +  Z(\hat g, + ; \hat g, - ) + Z(\hat g^2,-; \hat g, - ) +
Z(\hat g^3, + ; \hat g, - )\Biggr]
\ee
These are all $\tau\to \tau+1$ images of each other so if we are interested in
the massless states, corresponding to the coefficient of $q^0 \bar q^0$ we can
say this is just
\be
Z(\CH_{\hat g})\vert_{\rm massless} = \frac{1}{2}
\Biggl[  Z(1,-; \hat g, - )  \Biggr]_{q^0 \bar q^0}
\ee
Now, because  models A and B violate the  DTF condition \eqref{eq:NSV-Mod2}
the characteristic vector $W_g$ of \cite{Harvey:2017rko} is nonzero. Since
in the   $\hat g^2$ twisted
sector the momentum is shifted by $W_g$ and since  the oscillators are untwisted that
means the ground state must be massive. Thankfully, we don't have to compute the
$\hat g^2$-twisted sector partition function for massless states.
Moreover, the $\hat g^3 = \hat g^{-1}$ sector should give a second copy of the
massless states from the $\hat g$-twisted sector.
Therefore, we conclude that the number of massless twisted states is just:

\be\label{eq:NumBos}
 \# {\rm massless\ twisted\ real \ scalars} =
\Biggl[  Z(1,-; \hat g, - )  \Biggr]_{q^0 \bar q^0}
\ee

Now, it is easy to show that
\be
Z( 1,-;\hat g,-) =  \vert \CD(\Gamma^g)\vert^{-1/2} 2^{(n_- + \tilde n_-)/2} F_L(\tau) \overline{ F_R(\tau) }   \Theta_{(\Gamma^g)^\vee}(\tau)
\ee
where $\Theta$ is the theta function without shift vector (following the conventions of \cite{Harvey:2017rko}), while
\be
F_L(\tau) := \left(\frac{1}{\eta(\tau)}  \right)^{24-n_-}
\left( \frac{\eta(\tau)}{\eta( \tau/2)}\right)^{n_-}
\ee
\be
F_R(\tau) =   \left(\frac{1}{\eta(\tau)}  \right)^{8-\tilde n_-}
 \left( \frac{\vartheta_3}{\eta} \right)^{4-\tilde n_-/2}
  \left( \frac{\eta(\tau) }{\eta( \tau/2)} \right)^{\tilde n_-}
  \left( \frac{\vartheta_2}{\eta(\tau) }  \right)^{\tilde n_- / 2 }
\ee
In particular we have the leading $q$-expansions:
\be
F_L = q^{-1 + \frac{n_-}{16} } \frac{1}{(1-q^{1/2})^{n_-} }  [ 1+ \CO(q) ]
\ee
\be
F_R = 2^{\tilde n_-/2}  q^{-\half + \frac{\tilde n_-}{8} }   [ 1+ \CO(q^{1/2} ) ]
\ee
Let us stress that our partition function is in the NS sector.
So \eqref{eq:NumBos} counts the number of real massless bosonic fields. They will form hypermultiplets with
fermions from the R sector. If we let $TW$ denote the number of massless twisted
real scalars then altogether the number of massless hypermultiplets is
\be
\# HM = \frac{1}{4} TW + n_-
\ee

Now for model B we have $n_-=16$ and $\tilde n_- = 4$. So $F_L$ starts at $q^0$ and
$F_R$ starts at $\bar q^0$ so there no contributions from the lattice theta function.
So
\be
\begin{split}
TW:= \# {\rm massless\ twisted\ real \ scalars\ for\ model\ B}  & =
 \vert\CD(\Gamma^{g_B})\vert^{-1/2} 2^{(n_- + \tilde n_-)/2}2^{ \tilde n_-/2}\\
 &   = 2^{12} \vert \CD(\Gamma^{g_B})\vert^{-1/2}\\
\end{split}
\ee
Using the method described below in section \ref{subsubsec:DiscriminantGroups} we find
$\vert\CD(\Gamma^{g_B})\vert = 2^8$ for both cases $Q^{222}$ and $Q^{224}$
and hence, plugging in the numbers there are $7$ vectormultiplets and $80$ hypermultiplets
for both choices of $\fF_L$.
Given the very different symmetry groups in these two cases we expect the dual Calabi-Yau threefolds to
be very different in spite of their common Hodge numbers.

Similarly, for model A we have $n_-=8$ so
\be
F_L = q^{-1/2} ( 1+ 8 q^{1/2} + \CO(q) ).
\ee
Now we can get a possible contribution from the theta function. Let $\CN_{1,0}$ be the set of
vectors $p$ in $(\Gamma^g)^\vee$ with $p_L^2 =1 $ and $p_R^2 =0 $.
Then
\be
\begin{split}
TW:= \# {\rm massless\ twisted\ real \ scalars\ for\ model\ A}  & =
 \vert \CD(\Gamma^{g_A})\vert^{-1/2} 2^{(n_- + \tilde n_-)/2}2^{ \tilde n_-/2} (8 + \vert \CN_{1,0}\vert )\\
 &= 2^{8} \vert \CD(\Gamma^{g_A})\vert^{-1/2}(8 + \vert \CN_{1,0}\vert ) \\
\end{split}
\ee
Using the method described below in section \ref{subsubsec:DiscriminantGroups} we find
$\vert\CD(\Gamma^{g_A})\vert = 2^{12}$ for $Q^{222}$ and
$\vert\CD(\Gamma^{g_A})\vert = 2^{10}$ for $Q^{224}$. Moreover $\vert \CN_{1,0}\vert = 72$
for $Q^{222}$ and  $\vert \CN_{1,0}\vert = 28$ for $Q^{224}$.  Hence, plugging in the numbers there are
$15$ vectormultiplets in both cases and $88$ hypermultiplets for $Q^{222}$ and $80$ hypermultiplets
for $Q^{224}$.

\subsection{The Fate Of Massless Charged Hypermultiplets}\label{subsubsec:MasslessCharged}

For applications to heterotic/type II duality we must work a little harder in the case of $Q^{222}$ and
$Q^{224}$. Indeed the numbers $88$ and $80$ of massless hypermultiplets is a little misleading because
some of those hypermultiplets
(those  corresponding to states with  $p\in \CN_{1,0}$)
 will be charged under the $16$ $U(1)$ gauge symmetries. If there is a
type II dual of the heterotic theory with zero vacuum expectation value for these charged scalar fields
then we would expect
the Calabi-Yau threefold to be somewhat singular. As in   \cite{Greene:1995hu,Kachru:1995wm}, we
expect that if the charged scalars take generic vacuum expectation values then the dual Calabi-Yau manifold will be
smooth(er).

String perturbation theory should produce some potential energy in the Lagrangian for these scalars consistent
with the general structure of $d=4$, $N=2$ supergravity. (See \cite{FreedmanVanProeyen}
chapters 20 and 21, or  \cite{deWit:2001bk}, for the general form such a potential
energy function can take.) If the massless charged hypers get vacuum expectation values they will Higgs  some collection of those left-moving $U(1)$ gauge
bosons corresponding to the states listed in \eqref{eq:UT-3} below. We will assume that
the potential for these charged scalar fields is sufficiently generic that all of the modes of the
charged scalars that are not Goldstone bosons and that can become massive in a way consistent with 
$N=2$ supersymmetry, will become massive. Of course, the Goldstone bosons get eaten.
\footnote{In principle this assumption could be checked by computing the correlation
functions of the vertex operators of the twisted charged massless scalar fields.
Such a computation goes way beyond the scope of this paper and we will simply assume
the potential energy generated by conformal perturbation theory is generic.}
Moreover, we will assume the vacuum is at a generic point in the moduli space of vacua
and that the situation is sufficiently generic that the number of broken generators
of the $U(1)^{24-n_-}$ gauge group is the number, $\CL_{1,0}$,
of linearly independent vectors in $\CN_{1,0}$.
Of course, such vacua are  outside the moduli space of toroidal orbifolds.

Given the above genericity assumptions we can say that, after Higgsing, the number of
vectormultiplets and massless neutral hypermultiplets is altered, for model A, to
\be
\begin{split}
\# {\rm vectormultiplets} & = 23-n_- - \CL_{1,0} = 15- \CL_{1,0}  \\
\# {\rm hypermultiplets} & = n_- +  2^{9} \vert \CD(\Gamma^{g_A})\vert^{-1/2}   + \frac{1}{4}\dim_{\IR}\CM_{\rm Higgs} \\
\end{split}
\ee
To compute the dimension of the Higgs branch we proceed naively and count degrees of freedom and subtract
the numbers of equations and symmetries to get
\footnote{To do this more properly one should  use equations $(21.16)$ and $(21.30)$ of
\cite{FreedmanVanProeyen}. Again this goes beyond the scope of this paper. As a naive guess we expect that the $U(1)$
symmetry of the graviphoton remains unbroken so that $N=2$ supersymmetry is unbroken. (If this assumption were false it
would be quite interesting.) Moreover, at weak string coupling one expects that the much simpler hyperk\"ahler quotient
construction appropriate to field theory can be used. (Certainly, this was assumed in \cite{Greene:1995hu,Kachru:1995wm}.)
Since $\CL_{1,0} < 24-n_-$ only a proper subgroup of the abelian gauge group can be Higgsed. We can write a finite
cover of the gauge group in the form $U(1)^{\CL_{1,0}} \times U(1)^{\rm unb}$ where $U(1)^{\rm unb}$ acts trivially
on the scalars. Then we take the hyperk\"ahler quotient with respect to $U(1)^{\CL_{1,0}}$. This might leave an
unbroken finite group gauge symmetry. }
\be
\dim_{\IR}\CM_{\rm Higgs} = \frac{2^{8}\cdot \vert \CN_{1,0} \vert}{ \vert \CD(\Gamma^{g_A})\vert^{1/2}}  - 4\CL_{1,0}
= \begin{cases} 4\cdot 60 & Q^{222} \\
4\cdot 42 & Q^{224} \\
\end{cases}
\ee
and hence
\be
\begin{split}
\# {\rm massless~ neutral ~ hypermultiplets} & =
\begin{cases} 76 & Q^{222} \\   66 & Q^{224} \\ \end{cases}   \\
\end{split}
\ee
%

\subsection{Computation Of Discriminant Groups, $\CN_{1,0}$ And $\CL_{1,0}$ }\label{subsubsec:DiscriminantGroups}

Clearly, to proceed we need to be able to compute the lattice $(\Gamma^g)^\vee$ both
for the order of the discriminant group $\CD(\Gamma^g)$ and in order to compute the
vectors enumerated by $\CN_{1,0}$. This is not at all easy, and is probably out of
reach using purely human computational methods. We proceed using
 the following method which can be easily implemented using a (sufficiently bright) computer
algebra system. Let $g$ be an automorphism of order two of a lattice $\Gamma$ and let
$\Gamma^g$ be the sublattice of $\Gamma$ consisting of vectors in $\Gamma$ which are invariant under
the action of $g$. It is not easy to compute $\Gamma^g$ directly, but given an explicit action of $g$
we can easily compute the lattice $(1+g) \Gamma $ consisting of vectors $(1+g)v$ with $v \in \Gamma$
as well as the lattice $P_g \Gamma$ with $P_g=(1+g)/2$.
The lattice $\Gamma^g$ is an overlattice of $(1+g) \Gamma$, meaning that $(1+g) \Gamma$ and $\Gamma^g$ have
the same rank and that there is an embedding $(1+g) \Gamma \hookrightarrow \Gamma^g$. This is clear since
every vector in $(1+g) \Gamma$ is invariant under $g$ and hence in $\Gamma^g$. However there might be vectors
in $\Gamma^g$ which are not in $(1+g) \Gamma$.  Similarly $P_g \Gamma$ is an overlattice of $\Gamma^g$, so we have a lattice
sandwich:
\be
(1+g) \Gamma \subset \Gamma^g \subset \frac{1}{2}(1+g) \Gamma \, .
\ee

\noindent It might be useful to keep two simple and illustrative examples in mind.

{\it Example 1.} Let $\Gamma$ by the $A_2$ root lattice with simple roots $r_1$, $r_2$ and let $g$ be the reflection
in the hyperplane orthogonal to $r_1$ so that $g: r_1 \rightarrow -r_1, r_2 \rightarrow r_1+r_2$. Vectors in $\Gamma^g$ are all multiples of $(1+g) r_2$. Thus in this example $\Gamma^g=(1+g) \Gamma = \IZ(r_1+2 r_2)$ while $P_g \Gamma = \IZ(\frac{1}{2} r_1 + r_2)$.

{\it Example 2.} Let $\Gamma$ be the two-dimensional square lattice with basis vectors that are unit vectors $e_x$, $e_y$
and let $g$ be a reflection in the $y$ axis. Then $\Gamma^g=P_g \Gamma = \IZ e_y$ while $(1+g) \Gamma= 2 \IZ e_y$.

To determine $\Gamma^g$ we now proceed as follows.
Let $r$ be the rank of $\Gamma$ and $s \le r$ the rank of $(1+g) \Gamma$. Let $b_a$, $a=1, \dots, r$ be a basis for $\Gamma$
and $u_j$, $j=1, \dots,  s$ a basis for $(1+g) \Gamma$. A basis $u_j$ can be obtained by applying the LLL lattice basis reduction algorithm \cite{LLL}
to the matrix formed by the vectors $(1+g)b_a$.
%
%
 We can clearly express the $u_j$ as integer linear combinations
of the $b_a$,
\be
u_j = \sum_a M_{j,a} b_a
\ee
for some $M \in Mat(s,r,\IZ)$. We can determine $M$ as follows. Taking the inner product on both sides
with $b_{a'}$ gives
\be
(u_j, b_{a'}) = \sum_a M_{j,a} (b_a, b_{a'}) = \sum_a M_{j,a} Q_{a,a'}
\ee
and since $Q_{a,a'}=(b_a,b_{a'})$ is invertible we can just solve for $M$
as
\be
M_{j,a} = (u_j, b_{a'}) Q^{-1}_{a',a}
\ee
Next, $u_j/2$ is a basis for $P_g \Gamma$ and any vector in
$\Gamma^g$ is in $P_g \Gamma$. However not all vectors in $P_g \Gamma$ are in $\Gamma^g$ because not all of
them are in $\Gamma$. A vector $\half \sum_j n_j u_j \in P_g \Gamma$ is in $\Gamma^g$ precisely when it is in $\Gamma$,
that is when $\half \sum_{j,a} n_j M_{j,a} b_a \in \Gamma$, and this in turn is equivalent to the set of
equations:
\be\label{gamgmod}
\sum_j n_j M_{j,a} =0 ~{\rm mod}~2   \qquad \forall a =1, \dots, r,
\ee
since elements of $\Gamma$ must be integer linear combinations of basis vectors.
Taking all $n_j$ even always gives a solution.  The nontrivial solutions, indicating that there are vectors in $\Gamma^g$ which
are not in $(1+g) \Gamma$, arise when some of the $n_j$ are odd integers.  To obtain $\Gamma^g$ we thus take the basis $u_j$ for
$(1+g) \Gamma$ and construct an over complete set of vectors by adjoining vectors corresponding to all the nontrivial solutions of
\eqref{gamgmod}. This set of vectors forms an overcomplete set of vectors for  $\Gamma^g$ and from this a basis can be constructed
using the LLL algorithm \cite{LLL}. Once one has a basis it is straightforward to compute the order of the discriminant group
from the determinant of the Gram matrix.

 One can similarly use the inverse Gram matrix to produce a basis for
$(\Gamma^g)^\vee$ and then, with some more effort one can solve for the vectors $p$ with $p_L^2=1$ and $p_R^2=0$
and check how many such vectors are linearly independent. We provide some details and intermediate steps of this
calculation for models based on $Q^{224}$ in Appendix \ref{App:Details-224}.
%
%

\subsection{The Discrete Symmetries}

Finally, recall that we are particularly interested in these models because they have large and interesting discrete
symmetries. These symmetries will include (a possible lift of) the left-moving crystal symmetry originating from
the original Conway group symmetry. (This is the left-moving part of the group denoted $F(\Gamma)$ in
\cite{Harvey:2017rko}.)  As noted above, the symmetry group $\Aut(\fF_L^\perp) \subset Co_0$ (which is, in general,
only a subgroup of the full group of automorphisms of $\fF_L^\perp$) is $G^{222}= U_6(2)$ and the subgroup of the
monomial group   $G^{224}= 2^{10}.M_{22}$.  Since we have chosen any shift vector to be right-moving we simply need to compute the centralizer of the
involutions $g_A$, $g_B$ and $g_C$.  The generators of $\Aut(\fF_L^\perp) \subset Co_0$ are provided in \cite{HM} and we can then use GAP and Magma to identify
the conjugacy class of the involutions $g_A$, $g_B$, and $g_C$ and then compute their centralizers in  $\Aut(\fF_L^\perp) \subset Co_0$.
By studying the normal subgroups of the centralizer groups we obtain the descriptions of the groups
given in Table \ref{table:HM222} and Table \ref{table:HM224}.

\subsection{Summary Of The Six Models}

We can summarize our results in two tables. The first table concerns HM222 and the second HM224.

\begin{table}
\begin{center}
\begin{tabular}{|c|c|c|c|}
\hline
H\"ohn-Mason 222  & Model A  & Model B & Model C  \\
\hline
signature$(\Gamma^g)$ & $(14,2)$  & $(6,2)$  & $(10,2)$  \\
 \hline
 $\CD(\fF_L)$ &  $\IZ_2\oplus \IZ_2 \oplus \IZ_3$ & $\IZ_2\oplus \IZ_2 \oplus \IZ_3$  & $\IZ_2\oplus \IZ_2 \oplus \IZ_3$  \\
\hline
Left-moving crystal symmetry & $\IZ_2^4.\IZ_2^4.A_4.{\rm wr} \IZ_2 . \IZ_2$   & $\IZ_2 . \IZ_2^8. {\rm Sp}(4,\IZ_3)$  &
$\IZ_2 \times \left( \IZ_2^8. \IZ_3^2. Q_8\right)$  \\
 \hline
DTF? &  Yes & Yes &  Yes!   \\
\hline
$\vert \CD(\Gamma^g)\vert $ & $2^{12}$  & $2^8$  & x    \\
\hline
  $(\vert \CN_{1,0}\vert, \CL_{1,0})$ & $(72,12)$ & x  &   x  \\
\hline
(\# VM, \# HM) & $(15,88)$ & $(7,80)$  & $(11,12)$    \\
\hline
$(h^{1,1},h^{2,1})$ of hypothetical dual CY $\fX$ & $(3,75)$ & $(7,79)$ &   $(11,11)$   \\
\hline
\end{tabular}
\caption{Results for models A,B,C based on Gram matrix  \#222 of \cite{HM}. \label{table:HM222}}
\end{center}
\end{table}

\vskip1in

\begin{table}
\begin{center}
\begin{tabular}{|c|c|c|c|}
\hline
 H\"ohn-Mason 224 & Model A  & Model B & Model C  \\
\hline
signature$(\Gamma^g)$ & $(14,2)$  & $(6,2)$  & $(10,2)$  \\
 \hline
 $\CD(\fF_L)$ & $\IZ_4 \oplus \IZ_4$  &  $\IZ_4 \oplus \IZ_4$  &  $\IZ_4 \oplus \IZ_4$   \\
\hline
Left-moving crystal symmetry & $\IZ_2^8.(\IZ_2 \times \IZ_2 \times D_8). GL_2(\IZ_7)$   & $\IZ_2^{10}.(\IZ_2^4:A_6)$  & $2^x.M_{12}$  \\
 \hline
DTF? & Yes & Yes  & No    \\
\hline
$\vert \CD(\Gamma^g)\vert $ & $2^{10}$  & $2^8$  &   x  \\
\hline
  $(\vert \CN_{1,0}\vert, \CL_{1,0})$ & $ ( 28, 14)   $ & x  &   x  \\
\hline
(\# VM, \# HM) & $(15,80)$ & $(7,80)$  & $(11,12)$    \\
\hline
$(h^{1,1},h^{2,1})$ of hypothetical dual CY $\fX$ & $(1,65)$   & $(7,79)$ &   $(11,11)$  \\
\hline
\end{tabular}
\caption{Results for models A,B,C based on Gram matrix  \#224 of \cite{HM}. \label{table:HM224}}
\end{center}
\end{table}

In the above tables the notation ``x''  means we will not need to know the answer  (or the entry is not applicable),
while ``wr''  refers to a wreath product, $Sp(4,\IZ_3)$ is the symplectic group of rank two
over the field $\IZ_3$ and $Q_8$ is the quaternion group.
%
%
Moreover, in writing the Hodge numbers for model A in the last line of the table
 we have assumed a generic potential giving mass to all the
charged massless hypermultiplet scalars, as described in section \ref{subsubsec:MasslessCharged}.

\section{Implications Of Heterotic-Type II Duality}\label{sec:HetTypeII}

Thanks to heterotic-type II duality the above results
should have some implications
 for the type II string and for Calabi-Yau geometry. Our
remarks in this section are rather more speculative than what we have discussed
above. We hope they are positively provocative.

\subsection{Finding Type II Duals}

We begin with a few comments on the potential IIA Calabi-Yau 3-fold
duals to the heterotic models discussed above. It is not, \emph{a priori}
obvious that heterotic/type II duality should apply to asymmetric orbifolds
of heterotic string. Moreover, as far as we are aware, given a heterotic
model  on $T^2 \times K3$ with four-dimensional
 $N=2$ supersymmetry there is unfortunately no constructive procedure for
 identifying the dual type IIA Calabi-Yau 3-fold. We will assume such
 a dual exists and make a few simple remarks concerning its properties.

The gauge symmetry is abelian and this limits the kinds of singularities 
the dual can have. Moreover, the Calabi-Yau should be K3-fibered, by the
``adiabatic argument'' of \cite{Vafa:1995gm}. Moreover, the
untwisted hypermultiplet moduli include moduli that one can use
to give a six-dimensional limit where the size of the $T^2$ grows.
(This follows from a detailed investigation of the lattice invariant
 under $(g_R,g_L)$.) Therefore, according to \cite{Aspinwall:1996mn},
$\fX$ should be elliptically fibered.

Assuming the holonomy of the dual $\fX$ to be generic the
standard  rule is that the Hodge numbers of $\fX$ are related to the
four-dimensional $N=2$ supergravity by
\be
h^{1,1}(\fX) = \# VM \qquad \qquad h^{2,1}(\fX) = \# HM -1
\ee
Unfortunately there can be multiple Calabi-Yau's matching these Hodge numbers.
For example for case $C$ we have $(h^{1,1},h^{2,1}) = (11,11)$.
These are the Hodge numbers of the famous Borcea-Voisin 3-fold used
in the FHSV construction \cite{Ferrara:1995yx}. On the other hand,
section 7.2.3 of \cite{Bouchard:2007mf} gives another construction
of a CY 3-fold $\fX$ with the same Hodge numbers and $\pi_1(\fX) = \IZ_2$.
It is not obvious that these 3-folds are equivalent.
(See \cite{Donagi:2008xy} for further relevant information.) Moreover, 
as we have seen with 222B and 224B, the discrete symmetries in the two examples are 
very different, even though the Hodge numbers are identical, suggesting that the 
hypothetical dual CY 3-folds must be different. 

Using the tables for the Calabi-Yau Explorer of B. Jurke we can identify 
candidate CY 3-folds with the expected Hodge numbers appearing in our tables above:

\begin{enumerate}

\item $(3,75)$ is discussed in several places including \cite{Green:1987cr} \cite{Klemm:2004km}.

\item $(7,79)$ is discussed in  \cite{Braun:2011ux,Kreuzer:1995cd,Kreuzer:2000xy}.

\item $(11,11)$ is discussed in  \cite{Bouchard:2007mf,Donagi:2008xy,Ferrara:1995yx}.

\item $(1,65)$ is discussed in \cite{Batyrev:2008rp,Green:1987cr,Kapustka}.


\end{enumerate}

One can learn a great deal about the potential dual Calabi-Yau by performing
certain one-loop computations in the heterotic orbifold
\cite{Angelantonj:2013eja,Harvey:1995fq,Henningson:1996jz,Klemm:2005pd,Marino:1998pg}.
This includes
the intersection numbers along with enumerative invariants associated with
holomorphic curves in the Calabi-Yau. The intersection numbers are particularly useful for
identifying the dual. To obtain these the relevant one-loop integral (for threshhold corrections of
the gauge couplings $\left(\frac{1}{g^2}\right)^{ab}$ of the abelian gauge symmetries) takes the form
\be\label{eq:GeneralForm}
\int_{\CF} \frac{d^2\tau}{\tau_2} \left\{  \sum_{I}  F_I(\tau)
  \Theta_{\Gamma^g}\left(\alpha_I, \beta_I; (p_L^a p_L^b  - \frac{G^{ab}}{4\pi \tau_2} )    \right) - b^{ab}_{\rm gauge} \right\}.
\ee
Here $\CF$ is a fundamental domain for $PSL(2,\IZ)$ acting on the $\tau$-upper half-plane,
the sum on $I$ is a finite sum resulting from a sum over orbifold sectors and elements of the discriminant
group of $\Gamma^g$, while $F_I(\tau)$ are simple modular forms for congruence subgroups,
 $\alpha_I, \beta_I$ are characteristics for a Siegel-Narain theta function with an
insertion of $p_L^a p_L^b  - \frac{G^{ab}}{4\pi \tau_2}$, and finally $b^{ab}_{\rm gauge}$ is
the beta-function for the gauge couplings.      The intersection numbers can be
extracted from the ``degenerate-orbit'' contribution to the integral. The result is a concrete
expression for the polynomial $d_{ABC} y^A y^B y^C$, where $d_{ABC}$ are the intersection numbers of the
hypothetical Calabi-Yau and $y^A$
are special coordinates on the vectormultiplet moduli space \eqref{eq:VM-modulispace}. The expression that emerges is a
sum of rational expressions in $y^A$ (and is not at all obviously a polynomial). The sum
is obtained by choosing a finite-index embedding of $\Lambda_0 = K\oplus II^{2,2}(n)$ for some positive integer
$n$ and some positive definite lattice $K$ into $\Gamma^g$. The resulting sum is then
a finite sum over $I$, the elements of the quotient group $\Gamma^g/\Lambda_0$ and
vectors in $K$ of norm-square two.  We hope to present the details elsewhere.

\subsection{Auto-Equivalences Of The Derived Category}

The theorem of Gaberdiel, Hohenegger, and Volpato mentioned above
was interpreted by Huybrechts in terms of the group of
auto-equivalences of the derived category of a K3-surface
\cite{Huybrechts:2013iwa}. Identifying D-brane bound states with
cohomology classes of moduli spaces of objects in the derived category,
it is natural to interpret Huybrechts' theorem via heterotic
type II duality. In particular, the list of relevant groups
that can be identified as autoequivalences preserving the
symplectic and stability structure is precisely the list of
groups preserving rank four lattices in the H\"ohn-Mason tables.

Given  our orbifold CSS compactifications
and heterotic-type II duality it is natural to speculate that
there is a generalization of Huybrecht's theorem to Calabi-Yau
pencils of $K3$ surfaces. That is, we consider CY 3-folds $\fX$
with   $\pi: \fX \to \IP^1$ such that the generic fibers of
$\pi$ are smooth $K3$ surfaces.  In this case one can
define a triangulated subcategory  $D^b(\fX)_{\rm vert}$
of $D^b(\fX)$,  roughly speaking generated by coherent sheaves
supported on the fibers of $\pi$, and it should be possible to
define stability conditions $\sigma_{\rm vert}$
on $D^b(\fX)_{\rm vert}$  compatible with a stability  condition $\sigma$
on $D^b(\fX)$ as defined in
\cite{Aspinwall:2009isa,BridgelandStability}. For very general reasons the autoequivalences
of $D^b(\fX)$ preserving $\sigma$ form a finite group
\cite{BridgelandPC}. In view of heterotic/type II duality and
our considerations regarding CSS compactifications it is natural
to expect that autoequivalences of $D^b(\fX)_{\rm vert}$
preserving $\sigma_{\rm vert}$  should be related to subgroups of
the Conway group fixing sublattices of $\Lambda$ of rank two.
In particular, we would predict that for duals to the models
discussed in the previous sections the groups listed in the
two tables (or perhaps extensions thereof) are the groups of
autoequivalences. Of course, this is a highly nontrivial mathematical
claim, one which might be quite challenging to confirm or disprove.

Clearly, these are only meant to be preliminary remarks, and we leave a fuller
discussion of this idea to another occasion.

\subsection{Categories Of D-Branes }

The above considerations also raise two potential applications to
categories of branes.
\footnote{In both these remarks we are ignoring potential large IR gravitational
effects of low-dimensional compactifications. We can try to avoid this by a $g_{\rm string}\to 0$
limit, potentially limiting the applicability of these remarks to a ``geometric engineering''
subcategory of branes, such as that discussed in \cite{Chuang:2013wt}.}

To introduce the first question recall that there is a large mathematical literature
interpreting the bound states of D-branes on Calabi-Yau manifolds $\fX$
in terms of triangulated categories with stability condition. For the
$IIA$ string we find the derived category $D^b(\fX)$ and for the $IIB$
string the Fukaya category ${\rm Fuk}(\fX)$. Homological mirror symmetry
identifies  $D^b(\fX)$ with ${\rm Fuk}(\fY)$
where $\fY$ is the mirror of $\fX$ \cite{Kontsevich:1994dn,Aspinwall:2009isa}.
(More precisely it is an $A_\infty$ equivalence of ${\rm Fuk}(\fY)$
with an $A_\infty$ version of $D^b(\fX)$.)

Our first question, is then: Is there a similar mathematical discussion of D-brane
bound states on $S^1 \times \fX$?  Of course, the wrapped D-branes will
either sit at a point in $S^1$ or wrap the $S^1$. So, for example, if $\fX$ is a
3-fold then in the IIA string typical bound states will consist of $D6-D4-D2-D0$
bound states in $\fX$ localized at a point in $S^1$ and wrapping a holomorphic cycle
in $\fX$. If we only consider standard Lagrangian branes in
${\rm Fuk}(\fX)$ then, without loss of generality, we can take the other brane
to be a D4 wrapping  $S^1 \times L$ where $L$ is a Lagrangian subvariety of $\fX$.
Simple considerations of  the ground-state energies of  open string states
between these two kinds of branes suggest that they might form bound states.
\footnote{
Assuming the (possibly nontransverse) intersections are generic there will
be 4 Neumann-Dirichlet type open strings in a formulation based on
perturbative open string theory in the NSR formalism. As explained in
Chapter 13, vol. 2 of \cite{Polchinski:1998rq} at leading order supersymmetry will be
unbroken and the worldsheet ground state energy in the NS sector will be zero.
However, since these configurations are in a curved Calabi-Yau manifold there
can be perturbative corrections to the potential energy of deformation scalars
and it is entirely possible that these configurations sit at a local
 maximum in the potential energy. In such a case a small perturbation could
 lead to an instability that could settle on a nontrivial supersymmetric bound state.   }
Thus, mathematically one might ask if there exists an extension of the ``union'' of
 $D^b(\fX)$ and ${\rm Fuk}(\fX)$ that includes nontrivial morphisms between A- and B- branes.
 Since $S^1 \times \fX$
 is neither complex nor symplectic one might at first think that topological field
 theory methods are completely unsuitable for investigating these issues but that is
 not the case: At the self-dual radius of $S^1$ there is an   $N=3$ superalgebra
 with ${\rm Spin}(3)$ $R$-symmetry \cite{Dixon:1988ac} and so in fact
 one can define a variety of $A$- and $B$-models and $A$- and $B$-branes.

Our second question is motivated by the interpretation of CSS points in terms of
type II moduli. Following the discussion of \cite{Aspinwall:1994rg,Aspinwall:1996mn}
(see also Appendix C of \cite{Moore:1998pn})
it is clear that Conway subgroup symmetry of the D-brane system only arises when
suitable flat RR fields are turned on. Our second question then is: How do
flat RR fields affect bound states of D-branes, and can these effects be incorporated
into the above categorical description? As an extreme version of this we note that
the type II dual to the distinguished compactification based on $\Gamma_*^{24;8}$
must be  a compactification
on $K3 \times T^4$ with suitable flat $B$-fields and flat $RR$-fields
turned on. As an example of the odd things that can happen, if we replace the Leech lattice  in
$\Gamma_*^{24;8}$ by
the $E_8^3$ Niemeier lattice then the heterotic compactification
should have a type II dual with  $E_8^3$ gauge symmetry,
something which seems rather exotic in the framework of F-theory.

Once again, these remarks are extremely preliminary, and we leave further investigation
of these ideas for another time.

\appendix

\section{Computation Of The Untwisted Sector For The Four-Dimensional CSS Models}\label{app:UntwistedSector}

It is most convenient to work in the light-cone NSR formalism when deriving the
massless spectrum of a  $\IZ_2$ orbifold with generator $( \{ g_L; \delta \}, g_R)$ where
$\{ g_L; \delta \}$ is Seitz notation for the crystallographic symmetry
$x \to g_L \cdot x + \delta$.

We divide up the indices of the oscillators as follows:

\begin{enumerate}

\item $\alpha_n^{\mu}$, $\mu=2,3$. (Left-moving) Oscillations in noncompact spacetime orthogonal to the light cone. They have $g_L=+1$.

\item $\alpha_n^{i}$, $i=1,\dots, 22-n_-$. Oscillations in  compact directions with $g_L=+1$.

\item $\alpha_n^{a}$, $a=1,\dots, n_-$. Oscillations in  compact directions with $g_L=-1$.

\item $\tilde \psi_r^{\mu}$, $\mu=2,3$. (Right-moving) Oscillations in noncompact spacetime orthogonal to the light cone. They have $g_R=+1$.

\item $\tilde \psi_r^{i}  $, $i=4,5$. Oscillations in  compact directions with $g_R=+1$.

\item $\tilde \psi_r^{a}  $, $a=6,7,8,9$. Oscillations in  compact directions with $g_R=-1$.

\end{enumerate}

The only massless untwisted sector states have $P=0$. One might worry about states with $P=(p_L;p_R)$
with $\half p_L^2 = +1$ and $\half p_R^2 = \half$.  But such vectors would not be even and hence don't
exist. Therefore the only massless untwisted sector states have $P=0$. It is easy to list them:

\begin{enumerate}

\item  Metric, $B$-field, and dilaton:
\be\label{eq:UT-1}
\alpha_{-1}^\mu \vert 0 \rangle \otimes \tilde \psi_{-1/2}^\nu \vert 0 \rangle
\ee
Note that the dilaton and $B$-field are both real scalars.

\item Two abelian gauge fields from right-moving invariant subspace of $\IR^{6,22}$:
\be\label{eq:UT-2}
\alpha_{-1}^\mu \vert 0 \rangle \otimes \tilde \psi_{-1/2}^i \vert 0 \rangle
\ee

\item $22-n_-$ abelian gauge fields from left-moving invariant subspace of $\IR^{6,22}$:
\be\label{eq:UT-3}
\alpha_{-1}^i \vert 0 \rangle \otimes \tilde \psi_{-1/2}^\nu  \vert 0 \rangle
\ee

\item $2(22-n_-)$ real scalar fields from the invariant subspaces of $\IR^{6,22}$:
\be\label{eq:UT-4}
\alpha_{-1}^i \vert 0 \rangle \otimes \tilde \psi_{-1/2}^j  \vert 0 \rangle
\ee

\item $4n_-$ real scalar fields from the anti-invariant subspaces of $\IR^{6,22}$:
\be\label{eq:UT-5}
\alpha_{-1}^a \vert 0 \rangle \otimes \tilde \psi_{-1/2}^b  \vert 0 \rangle
\ee

\end{enumerate}

We get a total of $96$ states - independent of $n_-$.  This is confirmed by
writing partition functions implementing the orbifold and GSO projections.

We can now organize these states  into $d=4$ $N=2$ multiplets as follows:
Two states from \eqref{eq:UT-1} correspond to the graviton. The other two are scalars.
There are $24-n_-$ vector fields. One of them is the graviphoton. Therefore there are
$23-n_-$  $U(1)$ vectormultiplets. Therefore, there are $2(23-n_-) = 46 - 2n_-$ real vectormultiplet
scalars. On the other hand, the number of scalars is
\be
TW + 2 + 2(22-n_- ) + 4n_- = TW + 46 +2n_-
\ee
where $TW$ is the number of real massless scalar fields from the twisted sectors.
Subtracting the $46-2n_-$ vectormultiplet scalars we have
\be
  TW + 4n_-
\ee
hypermultiplet scalars, and therefore $\frac{1}{4}TW + n_-$ hypermultiplets.

\section{The Binary Golay Code, The Steiner System $S(5,8,24)$, And The Leech Lattice}\label{App:GolaySteiner}

We recall a few key definitions. The extended binary Golay code $\CG$ is a $12$-dimensional
subspace of $\IF_2^{24}$. The vectors in $\CG$, written as strings of $24$ zeroes and ones,
are called   codewords.
These codewords are divided up into $759$ \emph{octads}, consisting of codewords with
$8$ nonzero  and $16$ zero entries, $759$ octad complements, $2576$ \emph{dodecads}, consisting of
codewords with $12$ nonzero entries and $12$ zero entries, along with the origin $(0^{24})$, and the codeword
$(1^{24})$. To any codeword we can assign a subset of   the twentyfour element set $\Omega := \{1, \dots, 24\}$
using the positions  of the nonzero coordinates. These subsets of $\Omega$  are known
as $\CC$-sets. Using the definition
\be\label{eq:C-Set-Sum}
S_1 + S_2 := (S_1 - S_2) \amalg (S_2 - S_1)
\ee
the power set
of $\Omega$ becomes a $24$-dimensional vector space over $\IZ_2$, thus recovering $\CG$ from the
set of $\CC$-sets. The collection of $8$-element $\CC$-sets - also referred to as \emph{octads} -
form a collection of $8$-element subsets of $\Omega$, known as the \emph{Steiner system}
$S(5,8,24)$.  The Steiner system is distinguished by the property that any five element subset of $\Omega$
is a subset of a unique octad. The octads of the Golay code generate the entire
space $\CG$ using the operation $+$ defined in \eqref{eq:C-Set-Sum}. The Mathieu group $M_{24}$ is
the group of permutations acting as automorphisms of $\CG$, or, equivalently, as automorphisms
of $S(5,8,24)$.

The Golay code can be used to give a concrete construction of the Leech lattice as
a  lattice embedded  in $\IR^{24}$ as follows.  We take $\Lambda$ to be the set of
vectors $x:=(x_1, \dots, x_{24}) = c   (n_1, \dots , n_{24} )$
where $c=1/\sqrt{8}$, and  $n_i \in 2\IZ + \epsilon$ with $\epsilon=0$ or $\epsilon =1$ for all $i$.
We require moreover that $\sum_{i=1}^{24}  n_i = 4\epsilon ~\mod~8$,
and finally we require that for each $x\in \Lambda$, each of the four sets:
\be
\CC_a(x) := \{ i \vert n_i = a ~\mod~ 4 \} \subset \Omega
\ee
for $a=0,1,2,3$ is either the empty set or a $\CC$-set in the Golay code derived from the Steiner system $S(5,8,24)$.

An explicit basis for the Leech lattice  is
\be\label{eq:LeechBasis}
\begin{split}
v_{1} & = c\{4, -4, 0, 0, 0, 0, 0, 0, 0, 0, 0, 0, 0, 0, 0, 0, 0, 0, 0, 0, 0, 0, 0, 0\}\\
v_{2} & = c\{4, 4, 0, 0, 0, 0, 0, 0, 0, 0, 0, 0, 0, 0, 0, 0, 0, 0, 0, 0, 0, 0, 0, 0\}\\
v_{3} & = c\{4, 0, 4, 0, 0, 0, 0, 0, 0, 0, 0, 0, 0, 0, 0, 0, 0, 0, 0, 0, 0, 0, 0, 0\}\\
v_{4} & = c\{4, 0, 0, 4, 0, 0, 0, 0, 0, 0, 0, 0, 0, 0, 0, 0, 0, 0, 0, 0, 0, 0, 0, 0\}\\
v_{5} & = c\{4, 0, 0, 0, 4, 0, 0, 0, 0, 0, 0, 0, 0, 0, 0, 0, 0, 0, 0, 0, 0, 0, 0, 0\}\\
v_{6} & = c\{4, 0, 0, 0, 0, 4, 0, 0, 0, 0, 0, 0, 0, 0, 0, 0, 0, 0, 0, 0, 0, 0, 0, 0\}\\
v_{7} & = c\{4, 0, 0, 0, 0, 0, 4, 0, 0, 0, 0, 0, 0, 0, 0, 0, 0, 0, 0, 0, 0, 0, 0, 0\}\\
v_{8} & = c\{2, 2, 2, 2, 2, 2, 2, 2, 0, 0, 0, 0, 0, 0, 0, 0, 0, 0, 0, 0, 0, 0, 0, 0\}\\
v_{9} & = c\{4, 0, 0, 0, 0, 0, 0, 0, 4, 0, 0, 0, 0, 0, 0, 0, 0, 0, 0, 0, 0, 0, 0, 0\}\\
v_{10} & = c\{4, 0, 0, 0, 0, 0, 0, 0, 0, 4, 0, 0, 0, 0, 0, 0, 0, 0, 0, 0, 0, 0, 0, 0\}\\
v_{11} & = c\{4, 0, 0, 0, 0, 0, 0, 0, 0, 0, 4, 0, 0, 0, 0, 0, 0, 0, 0, 0, 0, 0, 0, 0\}\\
v_{12} & = c\{2, 2, 2, 2, 0, 0, 0, 0, 2, 2, 2, 2, 0, 0, 0, 0, 0, 0, 0, 0, 0, 0, 0, 0\}\\
v_{13} & = c\{4, 0, 0, 0, 0, 0, 0, 0, 0, 0, 0, 0, 4, 0, 0, 0, 0, 0, 0, 0, 0, 0, 0, 0\}\\
v_{14} & = c\{2, 2, 0, 0, 2, 2, 0, 0, 2, 2, 0, 0, 2, 2, 0, 0, 0, 0, 0, 0, 0, 0, 0, 0\}\\
v_{15} & = c\{2, 0, 2, 0, 2, 0, 2, 0, 2, 0, 2, 0, 2, 0, 2, 0, 0, 0, 0, 0, 0, 0, 0, 0\}\\
v_{16} & = c\{2, 0, 0, 2, 2, 0, 0, 2, 2, 0, 0, 2, 2, 0, 0, 2, 0, 0, 0, 0, 0, 0, 0, 0\}\\
v_{17} & = c\{4, 0, 0, 0, 0, 0, 0, 0, 0, 0, 0, 0, 0, 0, 0, 0, 4, 0, 0, 0, 0, 0, 0, 0\}\\
v_{18} & = c\{2, 0, 2, 0, 2, 0, 0, 2, 2, 2, 0, 0, 0, 0, 0, 0, 0, 0, 0, 0, 0, 0, 2, 2\}\\
v_{19} & = c\{2, 0, 0, 2, 2, 2, 0, 0, 2, 0, 2, 0, 0, 0, 0, 0, 0, 0, 0, 0, 0, 2, 0, 2\}\\
v_{20} & = c\{2, 2, 0, 0, 2, 0, 2, 0, 2, 0, 0, 2, 0, 0, 0, 0, 0, 0, 0, 0, 2, 0, 0, 2\}\\
v_{21} & = c\{0, 2, 2, 2, 2, 0, 0, 0, 2, 0, 0, 0, 2, 0, 0, 0, 0, 0, 0, 2, 0, 0, 0, 2\}\\
v_{22} & = c\{0, 0, 0, 0, 0, 0, 0, 0, 2, 2, 0, 0, 2, 2, 0, 0, 2, 2, 0, 0, 2, 2, 0, 0\}\\
v_{23} & = c\{0, 0, 0, 0, 0, 0, 0, 0, 2, 0, 2, 0, 2, 0, 2, 0, 2, 0, 2, 0, 2, 0, 2, 0\}\\
v_{24} & = c\{-3, 1, 1, 1, 1, 1, 1, 1, 1, 1, 1, 1, 1, 1, 1, 1, 1, 1, 1, 1, 1, 1, 1,1\}\\
\end{split}
\ee
This is almost identical to the basis given in \cite{nebeleech} except that we have taken a linear combination
of the first two basis vectors of \cite{nebeleech} so that all basis vectors have length squared four.
One may easily check that the Gram matrix is an even unimodular matrix.

A generating set for the $\CC$ sets of the Golay code can be read off from the above basis.
For each of the basis vectors $v_i$ only one of the four sets $\CC_a(v_i)$, $a=1,2,3,4$ is
nonempty. Let $S_i$ denote the corresponding $\CC$-set.   We will need:
\be
\begin{split}
S_8 & =  \{1,2,3,4,5,6,7,8 \}\\
S_{20} & =   \{ 1,2,5,7,9,12,21,24\}\\
S_{22} & =  \{9,10,13,14,17,18,21,22\}\\
S_{23} & = \{9,11,13,15,17,19,21,23\} \\
\end{split}
\ee
Two dodecads which are useful in our constructions are:
\be\label{eq:Dodecad}
\begin{split}
S_{20} +S_{22} & = \{ 1,2,5,7,10,12,13,14,17,18,22,24 \} \\
S_{8} + S_{12} + S_{22} & = \{ 5, 6,7, 8, 11,12,13,14,17,18,21,22 \} \\
\end{split}
\ee

We note that our choice of octads is slightly non-standard.
See the table of octads in \cite{Todd}. We also note that   there are more
efficient ways to generate them, as described in \cite{SPLAG}.

An important subgroup of ${\rm Co}_0$ for our considerations is the
``monomial subgroup,'' isomorphic to $2^{12}.M_{24}$. The action of this
group is easily described in terms of the above model for the Leech lattice.
$M_{24}$ is the group of permutations of coordinates
preserving the extended binary Golay code. Moreover, one can flip coordinates
on the $\CC$-sets of the extended binary Golay code. Let $\epsilon_{S}$ be
such a sign flip on a $\CC$-set $S$. Then clearly
\be
\epsilon_{S_1}\epsilon_{S_2} = \epsilon_{S_1+S_2}
\ee
and hence a set of generators for the subgroup of sign-flips is given by
$\{ \epsilon_{S_a} \}$ where $S_a$ runs over a basis for the Golay code.
Therefore the group of sign-flips is isomorphic to $\IZ_2^{12}$.

\section{Details On Lattice Computations: HM222}\label{App:Details-222}

\subsection{Construction Of $\Gamma^{22,6}$}

Recall from the discussion in \ref{subsec:LatticeLemma} that construction of an even self-dual lattice of signature $(22,6)$ starts with a pair of
isometric sublattices $\fF_L \subset \Gamma$ and $\fF_R \subset \Gamma_8$.

We first construct the lattice $\fF_R$ by choosing basis vectors
\begin{align}
\tilde u_1 &= u_1-u_2 = (0,0,0,0,1,1,1,-1) \\
\tilde u_2 &= u_2-u_3 = \frac{1}{2}(-1,1,1,-3,-1,-1,-1,1)
\end{align}
which are clearly in the span of the basis vectors for the Hadamard invariant sublattice $u_1, \cdots u_4$,have the desired Gram matrix
\be
Q^{222}= \begin{pmatrix} 4 & -2 \\ -2 & 4 \end{pmatrix} \, .
\ee
and generate a primitive sublattice of $\Gamma_8$.
We set $\fF_R = \IZ \tilde u_1 + \IZ \tilde u_2$.

The dual lattice $\fF_R^\vee$ has a basis $\tilde f_i= G^{ij} \tilde u_j$  which gives
\begin{align}
\tilde f_1&= \frac{1}{6} ( 2\tilde u_1 +  \tilde u_2 ) \\
\tilde f_2 &= \frac{1}{6} ( \tilde u_1 + 2 \tilde u_2 ) \, .
\end{align}
The generators of the discriminant group $\fF_R^\vee/\fF_R$ are as follows. Note that
\begin{align}
\tilde f_1+ \tilde f_2 &= \frac{1}{2}(\tilde u_1+\tilde u_2) \, ,\\
\tilde f_1- \tilde f_2 &= \frac{1}{6} (\tilde u_1-\tilde u_2) \, .
\end{align}
Thus we can take the generators of $\fF_R^\vee/\fF_R= \IZ_2 \oplus \IZ_2 \oplus \IZ_3$ to be
\begin{align}
\tilde c_1 &= \frac{1}{2}(\tilde u_1 + \tilde u_2) \, , \\
\tilde c_2 &= \frac{1}{2} \tilde u_1 \, , \\
\tilde c_3 &= \frac{1}{3} (\tilde u_1- \tilde u_2) \, .
\end{align}
Note that any  $f^* \in \fF_R^\vee$ is expressible as an integer linear combination of the $\tilde c_i$ and $2 \tilde c_1, 2 \tilde c_2, 3 \tilde c_3$ are in
$\fF_R$.

The lattice $\fF_R^\perp$ has basis vectors which may be chosen to be
\begin{align*}
\tilde v_1 &=  (1,0,1,0,0,0,0,0) \\
\tilde v_2 &= (0,1,-1,0,0,0,0,0) \\
\tilde v_3 &= (0,0,0,0,1,0,0,1) \\
\tilde v_4 &= (0,0,0,0,0,-1,0,-1) \\
\tilde v_5 &= \frac{1}{2} (1,-1,-1,-1,-1,1,-1,-1) \\
\tilde v_6 &= \frac{1}{2} (-1,1,1,1,-1,1,-1,-1)
\end{align*}
with Gram matrix
\be
G^{222}_{\fF_R^\perp}=\begin{pmatrix} 2 & -1 & 0 & 0 & 0 & 0 \\
                         -1 & 2 & 0 & 0 & 0 & 0 \\
                         0 & 0 & 2 & -1 & -1 & -1 \\
                         0 & 0 & -1 & 2 & 0 & 0  \\
                         0 & 0 & -1 & 0 & 2 & 0 \\
                         0 & 0 & -1 & 0 & 0 & 2
                         \end{pmatrix}
\ee
which we recognize as the Cartan matrix of $D_4 \oplus A_2$. Thus $\fF_R^\perp = \Lambda_{root}(D_4 \oplus A_2)$.
The dual lattice $(\fF_R^\perp)^\vee$ has a basis $\tilde g_i= (G^{222}_{\fF_R^\perp})^{ij} \tilde u_j$  which gives
\begin{align*}
\tilde g_1 &= \frac{1}{3} (2,1,1,0,0,0,0,0) \\
\tilde g_2 &= \frac{1}{3} (1,2,-1,0,0,0,0,0) \\
\tilde g_3 &=  (0,0,0,0,1,0,-1,0) \\
\tilde g_4 &=  \frac{1}{2} (0,0,0,0,1,-1,-1,-1)\\
\tilde g_5 &=  \frac{1}{4}(1,-1,-1,-1,1,1,-3,-1)\\
\tilde g_6 &= \frac{1}{4} (-1,1,1,1,1,1,-3,-1)  \, .
\end{align*}
It is not hard to see that one can choose the generators of the discriminant group $(\fF_R^\perp)^\vee / \fF_R^\perp \simeq \IZ_2 \oplus \IZ_2 \oplus \IZ_3$
to be
\begin{align*}
\tilde d_1 &= \tilde v_5  \, ,\\
\tilde d_2 &= \tilde v_6 \, , \\
\tilde d_3 &= \tilde v_1 \, .
\end{align*}
Since $(c_1^2,c_2^2,c_3^2)=(1,1,4/3)$ and $(d_1^2,d_2^2,d_3^2)=(1,1,2/3)$ we have $c_i^2 = -d_i^2 ~{\rm mod~}2$ for $i=1,2,3$ so the map
$(c_1,c_2,c_3) \rightarrow (d_1,d_2,d_3)$ gives an isometry $\CD_-(\fF_R) \cong \CD_+(\fF_R^\perp)$ as described in  subsection \ref{subsec:LatticeLemma}.

The construction of $\fF_L$ and $(\fF_L^\perp)^\vee$, the lattice $\Gamma$ and the invariant lattice $\Gamma^g$ as well as the
count of massless states in the twisted sector for HM22 follows very similar lines to the computations for HM224 for which details are
provided in the following section. As a result we omit them here and will provide them upon request.

\section{Details On Lattice Computations: HM224}\label{App:Details-224}

\subsection{Construction Of $\Gamma^{22,6}$}\label{App:Explicit-22-6}

Let us begin with the construction of $\fF_R$ and $\fF_R^\perp$. We take
$\fF_R$ to be generated by
\be\label{eq:FR-Gens}
\begin{split}
\tilde u_1 & =   (1,0,0,1,1,0,0,1) \\
\tilde u_2 &   = (1,1,1,-1,0,0,0,0)\\
\end{split}
\ee
These are fixed by the Hadamard involution, generate the required Gram matrix, and
generate a primitive sublattice.
It is straightforward to compute a basis for the orthogonal lattice $\fF_R^\perp$:
%
%
%
\be
\begin{split}
\tilde v_1 &= \{0,0,0,0,0,1,1,0 \} \\
\tilde v_2 &= (1/2) \{1,-1,-1,-1,1,-1,-1,-1 \} \\
\tilde v_3 &= \{0,0,0,0,-1,0,0,1 \} \\
\tilde v_4 &= \{0,1,-1,0,0,0,0,0,\} \\
\tilde v_5 &= (1/2) \{1,-1,1,1,-1,1,-1,-1 \} \\
\tilde v_6 &= \{0,0,0,0,0,-1,1,0 \}
\end{split}
\ee
The Gram matrix of the $\tilde v_i$ is
\be
\begin{pmatrix}
2 & -1 & 0 & 0 & 0 & 0 \\
-1 & 2 & -1 & 0 & 0 & 0 \\
0 & -1 & 2 & 0 & 0 & 0 \\
0 & 0 & 0 & 2 & -1 & 0 \\
0 & 0 & 0 & -1 & 2 & -1 \\
0 & 0 & 0 & 0 & -1 & 2
\end{pmatrix}
\ee
which we recognize as the Cartan matrix for $A_3 \oplus A_3$. This makes it clear that
the discriminant group $\CD(\fF_R^\perp)$ is $\IZ_4 \oplus \IZ_4 $.

By  inverting the Gram matrix of the $\tilde u_i$ we
can obtain explicit generators of $\CD(\fF_R)$
\be
\begin{split}
\tilde c_1 &= (1/4) \{1,0,0,1,1,0,0,1 \} \\
\tilde c_2 &= (1/4) \{1,1,1,-1,0,0,0,0 \}
\end{split}
\ee
and similarly we find explicit generators of $\CD(\fF_R^\perp)$:
\be
\begin{split}
\tilde d_1 &= (1/4) \{ 1,-1,-1,-1,0,2,2,0 \} \\
\tilde d_2 &= (1/4) \{1,2,-2,1,-1,0,0,-1 \} \, .
\end{split}
\ee

Since $\tilde c_1^2= \tilde c_2^2= 1/4$ while $\tilde d_1^2= \tilde d_2^2= 3/4$ and $\tilde d_1 \cdot \tilde d_2 =0$ we can construct
the isometry of discriminant groups  $\psi_R$ by mapping $ (\tilde c_1, \tilde c_2) \rightarrow (\tilde d_1 + 2 \tilde d_2, 2 \tilde d_1 + \tilde d_2)$.

%
%
%

On the left-moving side we need to proceed a little more indirectly to avoid having to
compute a basis for the $22$-dimensional sublattice $\fF_L^\perp \subset \Lambda$.
Now $\fF_L$ is spanned by $v_1, v_2$. Therefore $\CD_{+}(\fF_L)$ is just
$\IZ/4\IZ \oplus \IZ/4\IZ$ with $q(n_1,n_2)= \frac{n_1^2}{4} + \frac{n_2^2}{4} ~\mod~ 2\IZ$.
Define
\be
k_i := v_1\cdot v_i \qquad  1\leq i \leq 24
\ee
\be
\ell_i := v_2\cdot v_i \qquad  1\leq i \leq 24
\ee
 Now define vectors in $\fF_L^\perp \otimes \IQ$:
\be
\rho_a := - \frac{k_a}{4} v_1 - \frac{\ell_a}{4}v_2 + v_a \qquad 3 \leq a \leq 24
\ee
One can check that these span $(\fF_L^\perp)^\vee$. Moreover  the isomorphism
\be
\psi_L: \CD_-(\fF_L^\perp) \rightarrow \CD_+(\fF_L) \cong \IZ_4 \oplus \IZ_4
\ee
must take the form
\be
\rho_a \rightarrow (k_a ~\mod~ 4, \ell_a ~\mod~ 4)
\ee
Moreover
\be
\rho_a \cdot \rho_b = - \frac{k_a k_b}{4} - \frac{\ell_a \ell_b}{4} + v_a \cdot v_b
\ee
Therefore, if we use the quadratic function
\be
q(\bar v) := - v^2 ~\mod~ 2
\ee
we get an isometry of discriminant groups since we can form the integral linear combinations
\be
\gamma_a: = \rho_a + \rho_{24} \qquad a \in \Sigma_1:= \{3,4,5,6,7,9,10,11,13,17\}
\ee
with $(k,\ell) = (0,1)$ and
\be
\gamma_{ba} := \rho_b - (\rho_a + \rho_{24}) \qquad a \in \Sigma_1 \quad \& \quad b \in \Sigma_2 := \{ 15,16,18,19\}
\ee
 with $(k,\ell) = (1,0)$. We can now take any of the above $\gamma_a$ for the glue vector $d_1$ and
 any of $\gamma_{ba}$ for the glue vector $d_2$.

Thus, finally, we have explicitly constructed the CSS compactification:
\be
\Gamma^{22,6} := \amalg_{r=0,1,2,3} \amalg_{r' = 0,1,2,3} \biggl( \Gamma_0^{22,6} + r (d_1;\tilde d_1) + r' (d_2;\tilde d_2)
\biggr)
\ee
where $ \Gamma_0^{22,6}= (\fF_L^\perp;0) \oplus (0; \fF_R^\perp) $.

\subsection{Doomed To Fail?}

One can check that $g=(g_L^{A,B,C} ; \sigma_2)$ acts as the identity matrix modulo $\Gamma_0^{22,6}$ on all the
glue vectors. Therefore $g$ is an involutive automorphism of the lattice $\Gamma^{22,6}$.
Moreover, one can check for $g_L^A$ and $g_L^B$ there are vectors with $(p, g\cdot p)$ odd,
while for $g_L^C$ all inner products $(p,g \cdot p)$ are even. (Checking the above statements
involves some nontrivial computation - details are available upon request.) Based on the analysis in
\cite{Harvey:2017rko} we conclude that the lattice involutions for models A,B lift to symmetries of order four of the
lattice CFT while for model C the symmetry of the CFT remains order two.

\subsection{The Invariant Lattice $\Gamma^g$ And Its Discriminant Group}

We briefly describe the computation of $\Gamma^g$ for these models following the technique discussed in section 3.7. For model B we construct
the lattice $(1+g_B) \Gamma$ and then check that \eqref{gamgmod} has no non-trivial solutions which implies that $\Gamma^{g_B}= (1+g_B) \Gamma$. A set of basis vectors for $\Gamma^{g_B}$ consists of the vectors $(w_a,;\tilde w_a)$, $a=1,,8$ with
\be
\begin{split}
w_1 &= c \{0, 0, 0, 0, 0, 0, 0, 0, 0, 0, 0, 0, 0, 0, 0, 0, 0, 0, 0, 0, 0, 0, 0, 0\} \\
w_2 &= c \{0, 0, 0, 0, 0, 0, 0, 0, 0, 0, 0, 0, 0, 0, 0, 0, 0, 0, 0, 0, 0, 0, 0, 0 \} \\
w_3 &= c \{0, 0, 0, 4, 0, 0, 0, 0, 0, 0, 0, 0, 0, 0, 0, 0, 0, 0, 0, 0, 0, 0, 0, 0 \} \\
w_4 &= c \{0, 0, 2, 2, 2, 2, 2, 2, 0, 0, 0, 0, 0, 0, 0, 0, 0, 0, 0, 0, 0, 0, 0, 0 \} \\
w_5 &= c \{0, 0, 0, 0, 4, 0, 0, 0, 0, 0, 0, 0, 0, 0, 0, 0, 0, 0, 0, 0, 0, 0, 0, 0 \} \\
w_6 &= c \{0, 0, 0, 0, 0, 4, 0, 0, 0, 0, 0, 0, 0, 0, 0, 0, 0, 0, 0, 0, 0, 0, 0, 0 \} \\
w_7 &= c \{0, 0, 0, 0, 0, 0, 4, 0, 0, 0, 0, 0, 0, 0, 0, 0, 0, 0, 0, 0, 0, 0, 0, 0 \} \\
w_8 &= c \{0, 0, 0, 0, 0, 0, 0, 4, 0, 0, 0, 0, 0, 0, 0, 0, 0, 0, 0, 0, 0, 0, 0, 0 \} \\
 \tilde w_1 &= \{0, 0, 0, 0, 1, 1, 1, -1 \} \\
 \tilde w_2 &= \{1, 0, 0, 1, -1, 0, 0, -1 \} \\
 \tilde w_3 &= \{1/2, 0, 0, 1/2, 0, 1/2, 1/2, -1 \} \\
 \tilde w_4 &= \{0, 0, 0, 0, 1/2, 1/2, 1/2, -1/2 \} \\
 \tilde w_5 &= \{1/2, 0, 0, 1/2, 0, 1/2, 1/2, -1 \} \\
 \tilde w_6 &= \{1/2, 0, 0, 1/2, 0, 1/2, 1/2, -1 \} \\
 \tilde w_7 &= \{ 1/2, 0, 0, 1/2, 0, 1/2, 1/2, -1 \} \\
 \tilde w_8 &= \{ 1/2, 0, 0, 1/2, 0, 1/2, 1/2, -1 \}
 \end{split}
 \ee

For model $A$ on the other hand there are $63$ non-trivial solutions to \eqref{gamgmod} and to construct $\Gamma^{g_A}$ we produce an overcomplete
basis by adjoining the solutions to \eqref{gamgmod} to the basis vectors of $(1+g_A) \Gamma$ and then compute a basis from this over complete set.
We then find for the Gram matrix of $\Gamma^{g_A}$
\be
{\rm Gram}(\Gamma^{g_A}) =\left(
\begin{array}{cccccccccccccccc}
 -4 & 0 & -2 & -2 & -2 & -2 & -2 & -2 & 2 & -2 & 2 & 2 & 2 & -2 & -2 & 0 \\
 0 & -4 & -2 & 0 & 0 & 0 & -2 & -2 & 0 & -2 & 0 & 0 & 0 & -2 & -2 & 2 \\
 -2 & -2 & 0 & -1 & 0 & -1 & -2 & -2 & 0 & -2 & 0 & 0 & 2 & -2 & -2 & 2 \\
 -2 & 0 & -1 & 2 & 0 & 0 & -1 & 0 & 2 & 0 & 2 & 2 & 0 & -1 & 0 & -1 \\
 -2 & 0 & 0 & 0 & 2 & 0 & -1 & -1 & 0 & 0 & 0 & 0 & 2 & 0 & -1 & 1 \\
 -2 & 0 & -1 & 0 & 0 & 2 & 0 & -1 & 2 & -1 & 0 & 0 & 0 & 0 & 0 & -1 \\
 -2 & -2 & -2 & -1 & -1 & 0 & 0 & -2 & 1 & -2 & 1 & 0 & 0 & -2 & -2 & 1 \\
 -2 & -2 & -2 & 0 & -1 & -1 & -2 & 0 & 1 & -2 & 1 & 2 & 0 & -2 & -2 & 1 \\
 2 & 0 & 0 & 2 & 0 & 2 & 1 & 1 & 2 & 1 & 0 & 0 & -2 & 1 & 2 & -2 \\
 -2 & -2 & -2 & 0 & 0 & -1 & -2 & -2 & 1 & 0 & 2 & 1 & 1 & -2 & -2 & 1 \\
 2 & 0 & 0 & 2 & 0 & 0 & 1 & 1 & 0 & 2 & 2 & 0 & -2 & 0 & 1 & -1 \\
 2 & 0 & 0 & 2 & 0 & 0 & 0 & 2 & 0 & 1 & 0 & 2 & -2 & 1 & 1 & -1 \\
 2 & 0 & 2 & 0 & 2 & 0 & 0 & 0 & -2 & 1 & -2 & -2 & 2 & 1 & 1 & 1 \\
 -2 & -2 & -2 & -1 & 0 & 0 & -2 & -2 & 1 & -2 & 0 & 1 & 1 & 0 & -2 & 1 \\
 -2 & -2 & -2 & 0 & -1 & 0 & -2 & -2 & 2 & -2 & 1 & 1 & 1 & -2 & 0 & 0 \\
 0 & 2 & 2 & -1 & 1 & -1 & 1 & 1 & -2 & 1 & -1 & -1 & 1 & 1 & 0 & 2 \\
\end{array}
\right)
\ee
with determinant $2^{10}$.

\subsection{The Massless Charged Twisted Sector Ground States}

In model A there are massless charged states appearing in the twisted sector of the orbifold. To count these states
we need to compute the theta function of the dual of the invariant lattice, $(\Gamma^{g_A})^\vee$. In particular we want to
count states with $p_R=0$ and $p_L^2=1$. To do this we first project onto the sublattice of states with $p_R=0$ and compute
its Gram matrix. This leads to
\be
\left(
\begin{array}{cccccccccccccc}
 1 & -\frac{1}{2} & 0 & \frac{1}{2} & 0 & 0 & 0 & 0 & \frac{1}{2} & \frac{1}{2} & 0 & 0 & 0 & 0 \\
 -\frac{1}{2} & 2 & -1 & -1 & 0 & -\frac{1}{2} & 0 & 0 & -1 & -1 & 0 & \frac{1}{2} & -\frac{1}{2} & 0 \\
 0 & -1 & 2 & -1 & 0 & 0 & \frac{1}{2} & -\frac{1}{2} & 0 & 0 & -1 & -\frac{1}{2} & \frac{1}{2} & -\frac{1}{2} \\
 \frac{1}{2} & -1 & -1 & 4 & 0 & \frac{1}{2} & -1 & 0 & 2 & 2 & 2 & -\frac{1}{2} & -\frac{1}{2} & 1 \\
 0 & 0 & 0 & 0 & 1 & 0 & 0 & 0 & 0 & \frac{1}{2} & \frac{1}{2} & 0 & 0 & 0 \\
 0 & -\frac{1}{2} & 0 & \frac{1}{2} & 0 & 1 & 0 & 0 & \frac{1}{2} & 0 & \frac{1}{2} & 0 & 0 & 0 \\
 0 & 0 & \frac{1}{2} & -1 & 0 & 0 & 1 & 0 & -\frac{1}{2} & -\frac{1}{2} & -\frac{1}{2} & 0 & 0 & 0 \\
 0 & 0 & -\frac{1}{2} & 0 & 0 & 0 & 0 & 1 & -\frac{1}{2} & 0 & 0 & 0 & 0 & 0 \\
 \frac{1}{2} & -1 & 0 & 2 & 0 & \frac{1}{2} & -\frac{1}{2} & -\frac{1}{2} & 2 & 1 & 1 & 0 & 0 & \frac{1}{2} \\
 \frac{1}{2} & -1 & 0 & 2 & \frac{1}{2} & 0 & -\frac{1}{2} & 0 & 1 & 2 & 1 & -\frac{1}{2} & 0 & \frac{1}{2} \\
 0 & 0 & -1 & 2 & \frac{1}{2} & \frac{1}{2} & -\frac{1}{2} & 0 & 1 & 1 & 2 & 0 & -\frac{1}{2} & \frac{1}{2} \\
 0 & \frac{1}{2} & -\frac{1}{2} & -\frac{1}{2} & 0 & 0 & 0 & 0 & 0 & -\frac{1}{2} & 0 & 1 & 0 & 0 \\
 0 & -\frac{1}{2} & \frac{1}{2} & -\frac{1}{2} & 0 & 0 & 0 & 0 & 0 & 0 & -\frac{1}{2} & 0 & 1 & 0 \\
 0 & 0 & -\frac{1}{2} & 1 & 0 & 0 & 0 & 0 & \frac{1}{2} & \frac{1}{2} & \frac{1}{2} & 0 & 0 & 1 \\
\end{array}
\right) \, .
\ee
The theta function of the above Gram matrix is
\be
1+28 q^{1/2} + 2156 q + \cdots
\ee
which leads to the claim in Table 2 that there are $28$ massless states appearing in the twisted sector for $HM224$ model $A$.
Inspection (using Magma) of the actual short vectors (with $p_L^2=1$) shows that there are $14$ independent short vectors, hence ${\cal L}_{1,0}=14$.

\section{A Case Of Mistaken Moonshine}\label{App:NoMoonshine}

The space of DH states $\CV^{\rm BPS}$ in a toroidal heterotic string compactification
is graded by the Narain lattice and consists of right-moving ground states
and any left-moving state satisfying level-matching \cite{Dabholkar:1989jt}. Using light
cone quantization the space of physical BPS states as a vector space is
\be
\CV^{\rm BPS} = \oplus_{P\in \Gamma} \CD^{\rm BPS}_P \otimes \tilde \CR
\ee
where $\tilde \CR$ are the right-moving ground states of the 10-dimensional superstring.
They transform as $8_v \oplus 8_s$ under the ${\rm Spin}(8)$ automorphism of the
transverse space to the lightcone. We refer to the spaces $\CD^{\rm BPS}_P$
as the \emph{BPS degeneracy spaces}.  In  a light-cone
gauge formalism,  the BPS degeneracy spaces $\CD^{\rm BPS}_P$ can be identified with
the level $N=N_P$ subspace of a Fock space of $24$ chiral bosons
with
\be
N_P := 1 + \half( P_R^2 - P_L^2)
\ee

\bigskip
\noindent
\textbf{Remark}:
For $d=4$ heterotic/type II  duality states that the DH states can be identified
with the boundstates of D4-D2-D0 branes for the type IIA string compactified
on $\IM^{1,5} \times K3$. These are precisely the BPS states that were
investigated in \cite{Katz:2014uaa}, especially with regard to their
$\fs\fo(4) \cong \fs\fu(2) \oplus \fs\fu(2)$ quantum numbers. Physically,
those are the quantum numbers under a subgroup of the little group of a
corresponding
unitary representation of the Lorentz group $Spin_0(1,5)$. Thus, it is
possible to read the comments in this section, for $d=4$, as comments
about the $\fs\fo(4)$ representation content of these D-brane boundstates.
\bigskip
\bigskip

In what follows we will consider several Fock spaces generated by
chiral bosons valued in a $G$-module $V$ for various groups $G$.   Thus we
denote
\be\label{eq:Fock}
\CF_{q}V  := \sym^{\bullet}(q V \oplus q^2 V \oplus q^3 V \oplus \cdots )
= \sym^{\bullet}_{q}(V) \otimes \sym^{\bullet}_{q^2}(V) \otimes \sym^{\bullet}_{q^3}(V) \otimes \cdots
\ee
where $q$ keeps track of the level $N$, and
\be
\sym^\bullet_t(V) : = \oplus_{j=0}^\infty t^j \sym^j(V)
\ee
Note that if we consider  $\CF_{q}(V)$ as a $q$-expansion with coefficients
in the monoid of $G$-modules then we can define an inverse in the
space of $q$-expansions with coefficients in the \underline{representation ring} of $G$.
This will be a key idea in the discussion below.

\subsection{What Is Moonshine?}

The dimensions of the BPS degeneracy spaces $\CD^{\rm BPS}_P$
are naturally written as sums of dimensions of
irreducible representations of $O(24)$. When we consider the
compactification \eqref{eq:Spacetime-d} an $O(d)$ subgroup is
selected as a subgroup of the little group of a particle and
hence the
subgroup $O(24-d) \times O(d) \subset O(24)$ is distinguished.  Of course, the continuous
symmetry is broken by the Narain lattice, but
if one only studies BPS degeneracies and not,
for example, the algebra of BPS states (in the
heterotic theory) or auto-equivalences of the
derived category (in the type IIA theory)
then that breaking is not visible. All the
crystal symmetry groups $G_L$ in the CSS compactifications
are subgroups of $O(24-d)$, so it would be silly
to speak of ``Moonshine'' with respect to these groups.
Nevertheless, one can ask if there are discrete groups
\underline{not} in $O(24-d)$ that act on the spaces of
BPS states in a way compatible with modularity.
Moonshine is concerned with the latter phenomenon.

We will presently be more precise about the last sentence of
the previous paragraph,
but first a specific example will help clarify what we are speaking
about. Consider the case of $d=1$. From Table 1 of \cite{HM}
we see that the largest CSS symmetry for $d=1$ is ${\rm Co}_2$, the subgroup of $\Co0$ that
stabilizes a Leech vector of length-squared equal to four.
The space of BPS states, $\CV^{\rm BPS}$ as a representation of
${\rm Co}_2  \times O(1)$, can be obtained from the
Fock space
\be
\CF_q( V_{23}\otimes \textbf{T} \oplus \textbf{1}\otimes \textbf{S})
\ee
where $V_{23}$ is the $23$-dimensional representation of
${\rm Co}_2$ and $\textbf{T}$ and $\textbf{S}$ are the trivial
and sign representations of $O(1)$. The $q$-expansion is
\be\label{eq:Fock-d=1}
\begin{split}
\CF( V_{23}\otimes \textbf{T} \oplus \textbf{1}\otimes \textbf{S})
& 1 \oplus q \biggl[ V_{23}\otimes \textbf{T} \oplus \textbf{1}\otimes \textbf{S}\biggr]\\
& \oplus q^2 \Biggl\{
\biggl[ \sym^2(V_{23}) \oplus V_{23}\oplus \textbf{1} \biggr] \otimes \textbf{T} \oplus \biggl[ V_{23}\oplus \textbf{1} \biggr] \otimes \textbf{S} \Biggr\} \\
\oplus q^3
\Biggl\{
\biggl[  \sym^3(V_{23}) \oplus V_{23}^{\otimes 2}
&
\oplus 2\cdot V_{23}\oplus \textbf{1} \biggr] \otimes \textbf{T}
\oplus \biggl[   \sym^2(V_{23})\oplus 2\cdot V_{23}\oplus 2 \textbf{1} \biggr] \otimes \textbf{S}\Biggr\} \\
& \oplus \cdots \\
\end{split}
\ee
Indeed ${\rm Co}_2  \times O(1)$ will be an automorphism of the algebra of BPS
states. However, if we just study the BPS degeneracies $\CD^{\rm BPS}_P$ as
representations of ${\rm Co}_2  \times O(1)$ we might as well study them as
representations of $O(23) \times O(1)$ since   we can simply regard $V_{23}$ as
the vector representation of $O(23)$: If we just look at the degeneracies
 there is nothing special about the discrete group ${\rm Co}_2 \subset O(23)$.

However, we can ask if there is a non-manifest action of the larger group $\Co0$
 on the BPS degeneracy spaces $\CD^{\rm BPS}_P$. Note that $\Co0$ does
 \underline{not} have a nontrivial $23$-dimensional representation - so
there is no manifest $\Co0\times O(1)$ action on \eqref{eq:Fock-d=1}
(compatible with the pullback to ${\rm Co}_2 \times O(1)$).
One way to investigate this is to play the ``SumDimension game'' that
goes back to McKay's observations about the Monster and which was used to
such great effect in \cite{Eguchi:2010ej}. The rules of the
game in our context (for $d=1$) are:

\bigskip
\noindent
\textbf{SumDim1}:  List the dimensions of the irreducible representations of $\Co0$:
\be\label{eq:ConwayDimList}
{\rm Irrep}({\rm Co}_0) =
\{ \textbf{1}, \textbf{24},\textbf{276},\textbf{299},
\textbf{1771},\textbf{2024},\textbf{2576},\textbf{4576},\dots \}
\ee

\bigskip
\noindent
\textbf{SumDim2}:  Decompose the coefficients of $q^n \textbf{T}$ and
$q^n\textbf{S}$ into the simplest possible
nonnegative combinations of the integers \eqref{eq:ConwayDimList}. ``Simplest''
means ``using the fewest number of parts.''  Such minimal decompositions are in general
not unique. Choose one.

\bigskip
\noindent
\textbf{SumDim3}: Use ``simple'' virtual representations for ``small'' degeneracies.
These are typically associated with massless or exceptional representations. In
our $d=1$ example the ``small'' representations are the $23$-dimensional representations
in the coefficient of $q \textbf{T}$  and the coefficient of $q^2 \textbf{S}$.

As an illustration of rule two consider the coefficient of $\textbf{T}$ at
level $2$:
\be
\dim \biggl( \sym^2(V_{23}) \oplus V_{23}\oplus \textbf{1} \biggr) = 300
\ee
Referring to \eqref{eq:ConwayDimList} we can decompose
\be
300 = 299 +1
\ee
or, equally well,
\be
300 = 276 + 24
\ee
Either choice allows us to define a $\Co0$ action on that component of the space of
BPS states. Once we have made such a choice for all the massive levels we have
defined the Fock space as a $\Co0 \times O(1)$-module.
Of course, there are infinitely
many such choices and we must search for some principle that tells us which, if any,
are interesting.

The criterion we will use is modularity of the $\Co0 \times O(1)$ characters:
\emph{ We will say that the space of DH states
exhibits ``Moonshine for a group $G$'' if
that space of states has a $G$-module structure
with $G$ commuting with the level operator and such that
the graded character of $g\in G$ is a modular
form for $\Gamma_0(m)$ where $g$ has order $m$.}

\subsection{Virtual Representations}

It is easy to make Fock spaces of virtual representations
whose characters are modular. For example, returning to the
example of $d=1$ discussed above, if we replace
\be
V_{23} \to V_{24} - \textbf{1}
\ee
and interpret $V_{24}$ as the $24$-dimensional representation of $\Co0$
then the Fock space
\be\label{eq:VirtualFock-d=1}
\CF( (V_{24} - \textbf{1})\otimes \textbf{T} \oplus \textbf{1}\otimes \textbf{S}) =
\CF(V_{24}\otimes \textbf{T} ) \otimes \frac{\CF( \textbf{1}\otimes \textbf{S})}{\CF(\textbf{1}\otimes \textbf{T})}
\ee
regarded as a $q$-expansion with coefficients in the representation ring of
$\Co0 \times O(1)$ makes perfect sense. Moreover, its graded characters
\be
\Tr g q^{L_0-1}
\ee
will be modular forms for $\Gamma_0(m)$ if $g$ has order $m$.

The expression \eqref{eq:VirtualFock-d=1} is, \emph{a priori}, only a $q$-expansion
of virtual representations. However, it is possible to show that in fact,
with the exception of the coefficient of $\textbf{T}$ at level $1$ and
$\textbf{S}$ at level $2$ all the massive representations are in fact
\underline{positive} combinations of $\Co0$ irreps! But, dear reader,
we hasten to add that  this is a somewhat
silly form of Moonshine because we can also regard $V_{24}$ as the
vector representation of $O(24)$ and the same positivity holds
for \eqref{eq:VirtualFock-d=1} as a representation of $O(24)\times O(1)$:
Again, there is nothing   magical about the discrete group $\Co0\times O(1)\subset O(24)\times O(1) $.

There are two ways to prove this claim of positivity. One proof identifies the
extension of $O(23) \times O(1)$ symmetry to $O(24) \times O(1)$ symmetry
as the familiar fact that in light-cone gauge only the massless little group
acts linearly on the transverse oscillators and the extension to the generators
of the massive little group involve expressions that are nonlinear in oscillators.
In more detail, the argument goes as follows:

Consider  $(D-1)$ bosons  $\alpha^i$, $i=1,\dots, D-1$
and one extra boson $\beta$. We consider the Fock
space $\CF(V)$ based on the $D$-dimensional   representation
of $O(D-1) \times O(1)$:
\be
V = \textbf{(D-1)}\otimes\textbf{T} \oplus  \textbf{1}\otimes \textbf{S}
\ee
where $\textbf{T}$ is the trivial representation of $O(1)$ and $\textbf{S}$ is the sign
representation of $O(1)$.

Since $V$ is $D$-dimensional, if we forget about the $O(1)$ quantum number
of $\beta$ then the level $N$ subspace of $\CF(V)$ is clearly a representation
of $O(D)$, and under the natural inclusion of $O(D-1) \hookrightarrow O(D)$
it becomes the same $O(D-1)$ representation as the pullback
under $O(D-1) \hookrightarrow O(D-1) \times O(1)$.
A nontrivial fact - which is clear by thinking of the space $V$ as a transverse
space in a light-cone gauge formulation of string theory in $\IM^{1,D+1}$
(for $D \leq 24$)
is that for $N>1$,  $\CF^N(V)$ is in fact a representation of $O(D+1)$
such that, under the inclusion of $O(D) \hookrightarrow O(D+1)$ it pulls
back to the natural $O(D)$ representation noted above.

Thus, unlike the $O(D)$ action, there is an  $O(D+1)$ action  \underline{not} induced from a
linear action on $V$. We can now define an action by an $O(1)$ subgroup of $O(D+1)$
by counting the number of $\beta$ oscillators modulo two. Of course
$O(1) \cong \IZ_2$, viewed multiplicatively and we consider the subgroup
defined by identifying the nontrivial generator as
\be
\sigma = (-1)^{N_{\beta}}
\ee
The centralizer of this involution in $O(D+1)$ is a copy of $O(D)$,
and therefore $\CF^N(V)$ is a true representation of $O(D) \times O(1)$.

A second, more computational argument shows that in fact the positivity persists for all $D$,
not just $D\leq 24$.

\subsection{$d=4$: Mistaken Moonshine}

Let us now return to the the case of $d=4$, corresponding, via
heterotic/type II duality  to
the degeneracies studied in \cite{Katz:2014uaa}. We are
guaranteed $O(20) \times O(4)$ symmetry of BPS degeneracy spaces $\CD^{\rm BPS}_P$.
Let us ask if these spaces are, in any natural way, in fact representations
of $M_{24}\times O(4)$. This would be nontrivial since the smallest nontrivial
representation of $M_{24}$ is $23$-dimensional.

Given the surprising cancellations observed in the $d=1$ case
the first thing we might ask is whether the analogous Fock
space of virtual representations exhibits similar cancellations.
Since $M_{24}$ has a $23$-dimensional representation $V_{23}$
we consider
\be\label{eq:VirtualFock-d=4}
\CF_q\left( ( \textbf{23} - 3\textbf{1})\otimes\textbf{1} \oplus \textbf{1}\otimes V_4 ) \right)
= \CF_q\left( \textbf{23}\otimes \textbf{1} \right) \frac{
\CF_q\left( \textbf{1}\otimes V_4 \right)}{\CF_q\left( \textbf{1} \otimes\textbf{1}\right)^3}
\ee
where $V_4$ is the vector representation of $O(4)$.
Once again, the virtue of this construction
is that the character of elements of $M_{24}\times O(4)$ of finite order $m$ are guaranteed to
be modular forms for $\Gamma_0(m)$. In this sense, equation \eqref{eq:VirtualFock-d=4} exhibits
``Moonshine'' for $M_{24}\times O(4)$.

It is straightforward to expand \eqref{eq:VirtualFock-d=4}
in $q$. Let us for example consider the level $2$ degeneracy of
the singlet of $O(4)$. This degeneracy is $231$, which happens to be the dimension of an
irrep of $M_{24}$. Unfortunately, the explicit virtual representation of $M_{24}$
in \eqref{eq:VirtualFock-d=4}
turns out to be
\be
\begin{split}
S^2V - 2 V +1 & = V_{252} -  V_{23} + 2 \\
\end{split}
\ee
It indeed has net dimension $231$, but it is not a true representation.

One might still ask if, nevertheless, one of the solutions of the
SumDimension game nevertheless turns out to be modular. We believe that the answer
to this question is also negative. To show this we begin by playing
the SumDimension game at low levels. The dimensions of irreducible
representations of $M_{24}$ are
\be
\begin{split}
{\rm Irrep}(M_{24}) = &
\{ 1, 23, 45, 45, 231, 231, 252, 253, 483, 770, 770, 990, 990,\\
&  1035,
1035, 1035, 1265, 1771, 2024, 2277, 3312, 3520, 5313, 5544, 5796,
10395 \} \\
\end{split}
\ee
Next we study the possible
characters of involutions   in $M_{24}$ at low level.
Because a given degeneracy can have several possible ``simplest'' decompositions
as sums of dimensions of irreps of $M_{24}$
already at level four there are twelve possible characters for the
conjugacy classes of involutions of $M_{24}$ commonly denoted  $2A$ and $2B$. One
finds, for example
\be\label{eq:Z2A}
Z_{2A} =1/q  +  8 + 36 q + 144 q^2 + 282 q^3  + \cdots
\ee
and eleven other possibilities, all differing in the coefficient of $q^3$.
A similar story holds for  $Z_{2B}$. Now, just based on this meager information
it might seem impossible to rule out modularity for $\Gamma_0(2)$.
After all, we do not know the weight, the multiplier system, nor the
higher order terms in the $q$-expansion!

The trick to showing that $Z_{2A}$ and its cousins cannot be modular
is based on studying the behavior of the character in the vicinity
of  $\tau_0 := (1+\I)/2$. Recall that $\Gamma_0(2)$ is generated by $T$ and
$ST^2S$, the latter transformation acting as
\be
\tau' =    S   T^2   S\cdot \tau = \frac{\tau}{1-2\tau}
\ee
In particular, note that $ST^2S \cdot \tau_0 = \tau_0 -1$.
Since \eqref{eq:Z2A} is an expansion in integer powers of $q$
it is invariant under $T$. Thus, $\tau_0$ is effectively
a fixed point of $ST^2S$ and one   can determine the multiplier system from
the weight, as explained in Appendix \ref{App:MuliplierSystem}.
Moreover, if $\tau = \tau_0 + \delta \tau$ with $\delta\tau$ small
then
\be
\tau' = \tau_0 - 1 - \delta \tau + \CO(\delta\tau^2)
\ee
and so expanding the transformation formula in powers of $\delta\tau$
we find that the  weight can be deduced from
\be\label{eq:w-formula}
w = \lim_{\epsilon\to 0} \frac{1}{2\epsilon} \log \left\vert \frac{ Z(\tau_0 - \I \epsilon)}{Z(\tau_0 + \I \epsilon)} \right\vert
\ee
Finally,  for $\tau=\tau_0$ the numerical value of $q$ is
$q = \exp[2\pi \I \tau] = - e^{-\pi}=-0.0432139... $.
Since the generating function of the dimensions is rapidly
convergent for this value of $q$ we have every reason to
expect that the graded characters are also rapidly convergent  and therefore
we can compute $w$ to a good approximation just from the
first four terms given in \eqref{eq:Z2A} and its cousins.
The result of this computation is a non-half-integral
weight $w = - 8.4...$ in all cases the series is converging
to a complicated decimal expansion. (The value of $w$ for
the twelve cases all differ only in the third
significant figure.)  Therefore,  if $Z_{2A}$
is to be modular then it cannot have a half-integral weight.
\footnote{It is natural to ask if this trick can be used to
evaluate modularity of candidate series for $\Gamma_0(N)$ for
$N>2$ where the weight and multiplier system is unknown.
In this case the analogous point would be $\tau_0 = (1+\I)/N$
and $ST^N S \tau_0 = \tau_0 - 2/N$, so unless the candidate
function satisfies the rather unnatural condition of being
a series in powers of $q^{N/2}$ there is in fact no generalization.}
While it is possible to develop a theory of automorphic
forms for $SL(2,\IZ)$ of general complex weight,
 all Moonshine and conformal field theory
constructions of which we are aware involve forms of
half-integral weight.
It thus seems extremely unlikely that there is $M_{24}$
Moonshine in the degeneracies of perturbative heterotic
BPS states for the compactification $\IM^{1,5} \times T^4$.

Clearly, this argument is not completely rigorous
mathematically. One might, for example, doubt the accuracy
of our numerical expansion, or whether the series is 
really rapidly convergent, so that there is in fact a 
solution of the SumDimension game so that the weight in 
fact converges to $w=-8.5$. This (remote) possibility could
in principle be ruled out rigorously as follows: In this
case the multiplier system can be deduced using the
method of Appendix \ref{App:MuliplierSystem}. One can
then find generators of the relevant space of modular forms
and check if it is possible to have an expansion in integers
with the first four coefficients given by $Z_{2A}$. We have
not performed this exercise.

\section{Deducing The Multiplier System From The Weight}\label{App:MuliplierSystem}

\subsection{Multiplier Systems At Half-Integral Weight}

Suppose that $f(\tau)$ is a modular form of half-integral weight $w$ transforming
with  a multiplier system under some congruence subgroup $\Gamma$ of $PSL(2,\IZ)$.
Up to a phase $f(\gamma\cdot \tau)$  is $(c\tau +d)^w f(\tau)$.
Here we state our rules for defining the phase precisely.

Since $\eta(\tau)$ is nonvanishing on the upper half-plane,
$g(\tau) := f(\tau)/(\eta(\tau))^{2w}$ is perfectly well-defined and holomorphic,  and
$g(\gamma\cdot \tau)/g(\tau) $ for $\gamma\in \Gamma$ has modulus one. On the other hand, it is
meromorphic in $\tau$. Therefore it must be a \emph{constant} in $\tau$.
So we write
\be
\frac{g(\gamma\cdot \tau)}{g(\tau)} = \Phi(\gamma)
\ee
Now it is clear that since the LHS does not depend on $\tau$ we have
\be
\Phi(\gamma_1\gamma_2) = \Phi(\gamma_1)\Phi(\gamma_2)
\ee
so $\Phi$ must be a unitary character on $\Gamma$.

Now, define:
\begin{equation}\label{eq:ufun} u(\gamma,\tau):= \eta(\gamma\cdot \tau)/\eta(\tau) \end{equation}
 For  $\gamma=T^\ell$ we
have $u(T^\ell,\tau) = e^{2\pi i \ell/24}$. In general if
\begin{equation}
\gamma
 = \begin{pmatrix} a & b \cr  c & d \cr \end{pmatrix}
 \end{equation}
 with $c\not=0$ we can write:
\begin{equation}\label{eq:ufunii} u(\gamma,\tau) = \phi(\gamma) \bigl( -i ( \vert c \vert
\tau + {\rm sign}(c) d)\bigr)^{1/2}\qquad  \end{equation}
where $\phi(\gamma)$ is a $24^{th}$ root of 1 and we choose the
principal branch of the logarithm. Note that $u(\gamma,\tau)$ is a
cocycle: $u(\gamma_1\gamma_2,\tau) = u(\gamma_1,\gamma_2\tau)
u(\gamma_2,\tau)$ (indeed  it is also a coboundary). There are explicit
formulae for $\phi(\gamma)$ in textbooks on analytic number theory.

We interpret the transformation law for $f(\tau)$ with
half-integral weight to be:
\be
f(\gamma\cdot \tau) = (u(\gamma,\tau))^{2w} \Phi(\gamma) f(\tau)
\ee

\subsection{Deriving The Multiplier System From the Weight For $\Gamma_0(2)$}

Suppose we have a function $f(\tau)$ which is invariant
under $\bar T: \tau \to \tau +1$ and transforms with some half-integral
weight $w$ under $\bar S \bar T^2 \bar S$ and hence has weight $w$ with
multiplier system under $\bar \Gamma_0(2)$. We wish to show how to derive
the multiplier system given the weight.

We have $u(\bar T, \tau) = e^{2\pi\I/24}$, and hence
\be
\Phi(\bar T) = e^{- 4\pi w \I/24}
\ee
Next,
\be
\eta( \bar S \bar T^2 \bar S\cdot \tau) = \left(-\I (2\tau-1) \right)^{1/2}
e^{-2\pi \I/24} \eta(\tau)
\ee
Therefore the transformation law for $f$ under the   generator $\bar S \bar T^2 \bar S$ of $\bar\Gamma_0(2)$ is
\be
f( \bar S \bar T^2 \bar S\cdot \tau) = \biggl[ \left(-\I (2\tau-1) \right)^{1/2}
e^{-2\pi \I/24}\biggr]^{2w} \Phi(\bar S \bar T^2 \bar S) f(\tau)
\ee

Now note that if
\be
\tau' =  \bar S \bar T^2 \bar S\cdot \tau = \frac{\tau}{1-2\tau}
\ee
then for $\tau_0 := (1+\I)/2$ we have
\be
\tau_0' = \tau_0 -1
\ee
Therefore, provided $f(\tau_0)$ does not vanish,  we get the character on the other generator:
\be
\begin{split}
\Phi(\bar S \bar T^2 \bar S) & = e^{4\pi w\I /24}\\
\Phi(\bar T) & = e^{-4\pi w\I /24}\\
\end{split}
\ee
%
%
%

\end{document}